# Comparative Studies of Six Programming Languages

Zakaria Alomari  
Concordia University  
Montreal, Canada  
zakaria.alomari@gmail.com  

Oualid El Halimi  
Concordia University  
Montreal, Canada  
oualid.elhalimi@gmail.com  

Kaushik Sivaprasad  
Concordia University  
Montreal, Canada  
kaushik.sivaprasad@gmail.com  

Chitrang Pandit  
Concordia University  
Montreal, Canada  
pandit.ck@gmail.com

## Abstract
*Comparison of programming languages is a common topic of discussion among software engineers. Multiple programming languages are designed, specified, and implemented every year in order to keep up with the changing programming paradigms, hardware evolution, etc. In this paper we present a comparative study between six programming languages: C++, PHP, C#, Java, Python, VB ;  These languages are compared under the characteristics of reusability, reliability, portability, availability of compilers and tools, readability, efficiency, familiarity and expressiveness.*

## 1. Introduction:

Programming languages are fascinating and interesting field of study. Computer scientists tend to create new programming language. Thousand different languages have been created in the last few years. Some languages enjoy wide popularity and others introduce new features. Each language has its advantages and drawbacks. The present work provides a comparison of various properties, paradigms, and features used by a couple of popular programming languages: C++, PHP, C#, Java, Python, VB. With these variety of languages and their widespread use, software designer and programmers should to be aware of the benefits and drawbacks each language could bring to their software solution and be careful when they make rational decisions. These languages are compared under the characteristics of reusability, reliability, portability, availability of compilers and tools, readability, efficiency, familiarity and expressiveness. Other criteria like the programming effort, run time efficiency, memory consumption, and database connectivity are disclosed by implementing and running the same set of programs using all the languages under study.

**Related work**

In this comparative study work, relevant research  have been made based on trusted websites, research papers, scientific articles, and textbooks.

**Overview**

The rest of the paper is organized as follows. First we introduce the programming languages under study and we introduce their development history including their important contributors, versions, paradigms. Then we will address how each language is evaluated, typing strategies used, and how memory is managed. Then, we will highlight the strengths and weaknesses for each language as well as the applications domains in which these languages are mainly used. A couple of experiments will be conducted to compare efficiency, GUI development, and database connectivity followed by analysis and results interpretations. Moreover, we will compare language features and GUI development facilities

# 1. Programming Languages:

## 1.1 PHP

### 1.1.1 History & Versions:

PHP Language (Hypertext Pre-processor) is a server-side scripting language designed for web development. It is a powerful language that runs on more than 244 million websites. Ramsus Lerdorf first created PHP in 1994. He wrote a series of Perl scripts used to enhance his personal website. The main objective to create these scripts was to enhance his personal website and improve its performance. Another reason to create PHP was to monitor his online resume and related information. Later on, Ramsus start to develop C scripts to add the ability to work with web forms and access databases [O-1]. One year later, Ramsus combined PHP with his own Form interpreter and then released PHP 2.0 on June 8, 1995. Two years later, two programmers, Zeev Suraski and Andi Gutmans, released PHP/FI in 1997 by rebuilding the PHP's core system [O-2]. In 1998, these programmers ported many PHP/FI utility tools into PHP3, which was released in 1998 to support multiple platforms, web servers, and a wide number of databases. This PHP3 version was the first widely used version at that time. PHP4 was developed by the PHP development team and Zend technologies to improve speed and reliability over PHP3 and then released in May 2000 [O-4]. In fact, PHP4 has introduced new features to the language such as reference and Boolean support, COM support on Windows, output buffering, and expaned object oriented programming [O-4]. Finally, PHP5 was introduced in July 2004. Powered by Zend Engine II, PHP5 was an improvement version of PHP4 and included many features such as improved support of object-oriented programming and provided a well-defined and consistent PDO interface to access databases. Other important feature implemented in PHP5 is the backward compatibility with earlier version of PHP [O-5]. Late binding feature was added to later version of PHP5 (V5.3). Some other significant variants between PHP5 and PHP4 are:

- Introducing interfaces.

- Incorporate static method with the properties.

- Allow declaring classes as final

- Introduce three level of visibility: public, private, and protected.

- Add support of exceptions.

PHP takes most of its language from C, java, and Perl [O-2]. Similarly to java and C++, PHP has native Boolean types. Also, in terms of keyword and syntax, PHP has similar keywords to most high-level languages that follow the C style. Similarly to C language, PHP stores whole numbers in a platform dependant range: 32 or 64-signed integer [O-1]. PHP borrows the If, while loops and function returns style from C and Java. Moreover, it uses most C built-in data types. In fact, PHP has not influenced the development of any programming language at the present date [O-8]. However, there are successful and complex tools and software systems in the market that are mainly written in PHP as Joomla CMS [O-9] and Facebook [10].

### 1.1.2 Paradigms

Like C++, PHP5 promotes the Object oriented paradigm since you can deal with object and define classes [O-30]. For instance:

```php
<?php

  class ExampleClass{

      } ?>
```

PHP objects and classes have properties and methods. Also, classes have contractors that runs when an object is instantiated. Furthermore, we can use SELF and $this to reference properties and methods inside the class. Also, PHP supports polymorphism and inheritance similarly to C++ [O-31].

PHP is a reflection-oriented language paradigm. A PHP program is able to observe and modify the program execution at run time. PHP program is able to tell about its properties and methods and altering those members [O-32]. One example of PHP reflection could be summarized on the following example [O-32]:

```php
// Nettuts.php
class Nettuts {
  function publishNextArticle($editor) {
    $editor->setNextArticle('135523');
    $editor->publish();
  }
}
// Editor.php
class Editor {
  private $name;
  public $articleId;
  function __construct($name) {
    $this->name = $name;
  }
  public function setNextArticle($articleId) {
    $this->articleId = $articleId;
  }
  public function publish() {
    // publish logic goes here
    return true;
  }
}

// Manager.php
require_once './Editor.php';
require_once './Nettuts.php';

class Manager {
  function doJobFor(DateTime $date) {
    if ((new DateTime())->getTimestamp() > $date->getTimestamp()) {
      $editor = new Editor('John Doe');
      $nettuts = new Nettuts();
```

```
        $nettuts->publishNextArticle($editor);
    }
  }
}
```

As you illustrated in this code, at run time, Nettus used an editor object. At run time, PHP inspects the received object and checks that it implements the publish() and setNextArticle() methods.

### 1.1.3 Strengths/Weaknesses and Application Areas

There is no perfect programming language. Each comes with advantages and fallouts.

PHP code is flexible. However, it could make debugging and refactoring a nightmare. PHP can be used everywhere and run on all browsers on different platforms. It is considered to be the fastest [O-42] when compared to other programming languages. It is fast when it comes to connect to databases and fetch data. That's why it is used for important corporations to perform fast client/server transactions. PHP is also known for its easy syntax which enables it to be used in schools and universities and make possible for novice students to start to build their first scripting applications. PHP scripts can be integrated with any web file: HTML, JavaScript, XML, JSP, and so forth. So it can be integrated to most websites to implement specific business logic. PHP has a number of disadvantages that could be overcome with good design and development methods like security and error handling. Hence, the developer should enforce security code and statements that are more likely to throw exceptions to ensure security feature and error handling as well as privacy. i.e: bank online transactions [O-42].

### 1.1.4 Evaluation Strategy

**Compilation & execution**:

PHP scripts are interpreted in the web server. PHP script is first translated into machine language: object code. Then the linker links this file and sends the results back to the client machine [O-37].

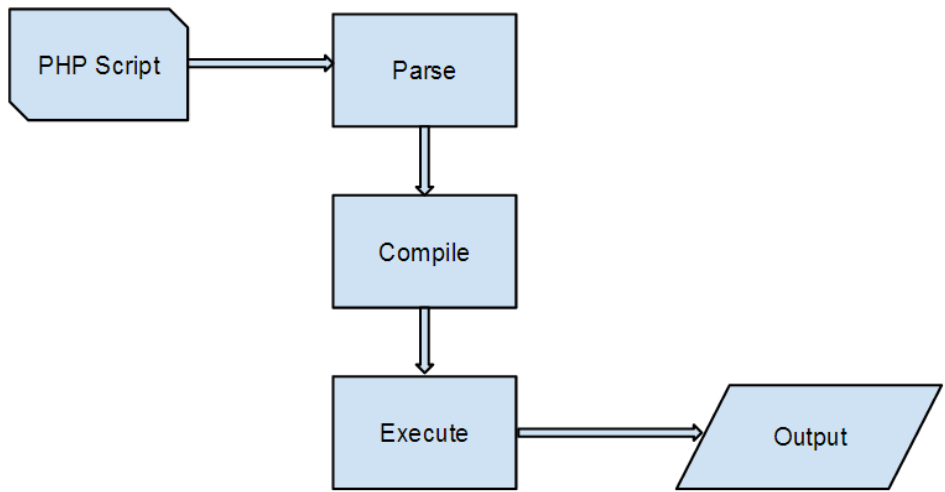

**Figure 1: Php program evaluation process**

Unlike a compiled language like C++ or C, PHP code is not compiled directly to machine code. In fact, it is compiled to an intermediary bytecode which is later interpreted by the PHP runtime engine. This evaluation process allows PHP to be easily, quickly and efficiently ported to a several different platforms [O-40].

Unlike C++, PHP gets evaluated by an interpreter. Both of them Convert a high level language into a binary form that is recognized by computer hardware [O-39]. Unlike C++ compiler, PHP interpreter translates one statement at a time and executes it while the Qt compiler translates the entire program at once and then executes it. As a matter of fact, C++ takes more time to get analyzed and processed comparatively to analyzing and processing php code. As a result, the overall execution time of C++ Compiled code is faster than the PHP interpreted code [O-40]

**Typing Strategies:**

PHP is a dynamically typed and weak language. PHP is dynamic since its variables must necessarily be defined before they are used. This implies that PHP does not require any explicit declaration of the variable before being used [O-33]. PHP is a weak-typing language since its variables are not a specific data type. Meaning: PHP variables are not bound to a specific data type. As an example [O-33]:

/* PHP code */

 <?php

$foo = "x";

A Comparative Studies of Programming Languages (Comparative Studies of Six Programming Language)

$foo = $foo + 2; // not an error  echo $foo;

?>

In this example, foo is first declared as a string type. Then, we add 2 to the string. This is allowed in PHP as a part of weak-type definition.

**Memory Management:**

Memory management is critical to PHP programs. The first reason to implement garbage collection mechanism for PHP is to reduce memory usage by releasing memory and cleaning up circular-referenced variables [O-6]. PHP garbage collection uses reference counting technique to manage its internal variables [O-6]. This technique is used to keep track of the sorted references to memory locations held by objects. When an object is destroyed, all references to this object are removed. However, reference counting technique fails to deal with address circular reference memory leak. Collecting Cycles mechanism handles this issue where memory clean up is invoked automatically when the root buffer is full or the gc_collect_cycles() is called [O-4].

However, running memory clean ups by PHP garbage collector could introduce performance decline and run time slowdowns when cycle collecting algorithm runs [O-4].

### 1.1.5 Available Compilers/IDEs

There are several available IDEs for both PHP and C++. Eclipse + PHPEclipse plugin is an open source add-on to the eclipse framework used to provide PHP tools [O-38]. It is available for Windows, Mac, and Linux operating systems. It is entirely developed in Java in order to add PHP support to eclipse IDE and combine HTML tags and PHP code in one file. It is a complete development tool for PHP that adds editing, formatting, syntax completion, outline view, running and debugging facilities to Eclipse IDE [O-39]. Eclipse supports PHP GUI development by allowing PHP integration with HTML. Web tools Platform comes with the Eclipse IDE for Java EE Developers [O-40] and has a web page HTML editor that let you see the elements you are editing in a preview window. WTP includes also HTML graphical editor and built-in applications to simplify web development. It also includes tools to support deploying, running, and application testing [O-41]. It can be downloaded from: http://www.eclipse.org/pdt/downloads/

Visual PHP is a complete PHP visual development environment designed for developers to quickly create PHP web applications and we services [O-42]. It is available for only Windows operating system. It contains a WYSIWYH HTML editor as well as several extensible modules that can be reused. The WYSIWYH implies a user interface that allows the user to drag and drop the GUI components to the UI and view the end result page at editing time [O-43]. It allows layout manipulation directly without dealing with code and without having to remember fields' names or layout commands. It can be downloaded from http://www.visual-php.com/en/

Adobe Dreamweaver is a web development tool developed originally by Macromedia in 1997 and then acquired by Adobe Systems in 2005 [O-44]. It provides a visual WYSIWYG editor with standard features as syntax highlighting, code completion, and real time type syntax checking to assist with code writing [O-45].

Its design view is built to make layout design and code generation as it allows developers to develop UI layouts using HTML. The tool has a built-in integrated browser to preview developed pages and also to allow content to be open locally. Like other HTML editors, Dreamweaver edits files locally and upload them to remote web server using FTP [O-46]. It is available for Windows, Mac OS, and Linux operating systems and can be downloaded from http://www.adobe.com/ca/products/dreamweaver.html

## 1.2 C++

### 1.2.1 History & versions:

C++ is a general purpose programming language. Designed by Bjarne Stroustrup in 1979. As a part of his PHD thesis, Stroustrup found that Simula owns important features that would help in building large software [O-11]. The main motivation was to adopt distributed computing in the UNIX operating system in the AT&T Bell Labs. Therefore, the idea to enhance the existing C language with Simula-like features came to mind. C was chosen thanks to the fact that it is a general-purpose language, fast, portable, and widely used [O-11]. In 1983, it was renamed to C++ as an increment to the C language. C++ added several features likes:

- Virtual functions.

- Operator overloading.

- Function name.

- References

- Constants

- Improved type checking

- One line comments support (e.g. beginning with //), to name a few.

In 1990 the C++ programing language was released and the first commercial implementation of C++ was implemented a few months later [O-12]. Other new features were introduced as multiple inheritance, abstract classes and static member functions, and const member functions [O-11]. In 1991, C++ language starts to evolve as well as the standard library. Therefore, other important features were added such as templates, exceptions, namespaces, and casting [O-12]. Then, in 1992, C++ standard libraries introduced a stream I/O library that provided facilities to replace the traditional C scanf and printf functions. C++ is standardized by the international Organization Standardization (ISO) and published in 2011.

C++ is considered to be a superset of C and it has helped in the development of many other programming languages like: Perl, PC, Lua, Pike, Ada95, PHP, D, C99, Flacon, and Seed7. For example, Java is derived from C++ but made simpler by getting rid of numerous C++ features like pointers, multiple inheritance, and operator overloading [O-13]. On the other hand, C# is also built from C++. It is built on the syntax and semantics of C++ and uses most of C++ features such as classes, parameter passing [O-16]. The Multiple inheritance feature was not adopted in either Java or Csharp due to the Diamond Problem [O-18]. The C++

A Comparative Studies of Programming Languages (Comparative Studies of Six Programming Language)

programming language was standardized initially in 1998 as ISO/IEC 4882:1998 and later amended by the 2003 technical corrigendum [O-11].

### 1.2.2 Paradigms

C++ is a multi-paradigm language: procedural, object oriented and generic [O-11]. C++ procedures are also called routines, methods, or functions and they define a computational to be carried out by the program [O-18]. C++ promotes procedural paradigm by providing facilities to pass arguments to functions and return values from them.

For example [O-22]:

```
double sqrt(double arg) {  … // code for calculating a square root }

void f() {   double root2 = sqrt(2);   // … }
```

This example illustrates the use of procedural programming in C++. Sqrt function contains a mathematical computation that will be used to set the root variable in the f function. In general, C++ functional paradigm emphasizes the use of functions that produce results depending on their input parameters and not in the program state [O-22]. In fact, functions are used to create order in a maze of algorithms using function calls and decide how the entire program will be executed [O-23].

C++ is an object oriented paradigm language. C++ represents concepts as objects and associated procedures called methods [O-24]. These C++ objects are usually class instances and mainly used to interact with each other to design computer programs and system applications. C++ is designed following the object oriented modularity approach [O-14] by defining classes, structs, objects, methods, and interfaces. Indeed, the basic block for object oriented paradigm in C++ is the class. This latter defines the data elements that define the state of the object and how these objects can be manipulated. Another C++ feature that promotes OO is encapsulation and information hiding by introducing interfaces and visibly access modifiers: public, private, and protected [O-19]. Furthermore, C++ promotes modularity by supporting inheritance, which makes it easy to maintain and reuse code, and extending components as needed by defining new subclasses with concrete implementations [O-19].

C++ is a generic programming paradigm. Generic paradigm mainly consists of abstracting concept so they will operate efficiently on any type that satisfies the interface being used by the algorithm. C++ promotes this paradigm by using templates [O-24]. C++ templates low functions and classes to operate with generic types. This feature allows a class or function to work on many different date types without the need to re-write the code for every single type [O-25].

### 1.2.3 Strengths and Weaknesses and application areas

C++ is a high level language that is easier to use than other forms of low level languages (i.e assembly). Though C++ takes much more space than low-level languages, it is easier to both learn and use and it is compatible with C code. It favours a mix of paradigms: Procedural, Object Oriented, Generic, and functional. It supports the feature of compile-once and run-everywhere which allows a common code base to be shared among different target hardware with different CPU and memory bounds [O-41]. C++ has a good support for generic programming and also supports classes, structs, and unions and allocates them into the heap. C++ programs have generally great performance compared to other high level languages like Java and CSharp [O-32]. Therefore, they are good candidates to develop operating systems, device drivers, embedded software thanks to the C-Embedded language, high performance client and server applications, and video games.

However, one major problem in C++ is memory allocation. There is no garbage collector and it is up to the programmer to clean up his mess after use. Another problem that arises is the goto instruction. This latter is bad for structured programming and breaks control flow [O-41]. Other C++ problems are: Multiple inheritance, member pointers, operator overloading and template syntax that is hard to read and maintain. These drawbacks put more effort on the software designer and developer to deal with these issues in the code to overcome these weaknesses. For instance, deal with memory management.

### 1.2.4 Evaluation Strategy
**Compilation and Execution**

C++ compilation process consists of four steps as the following figure illustrates [O-35]:

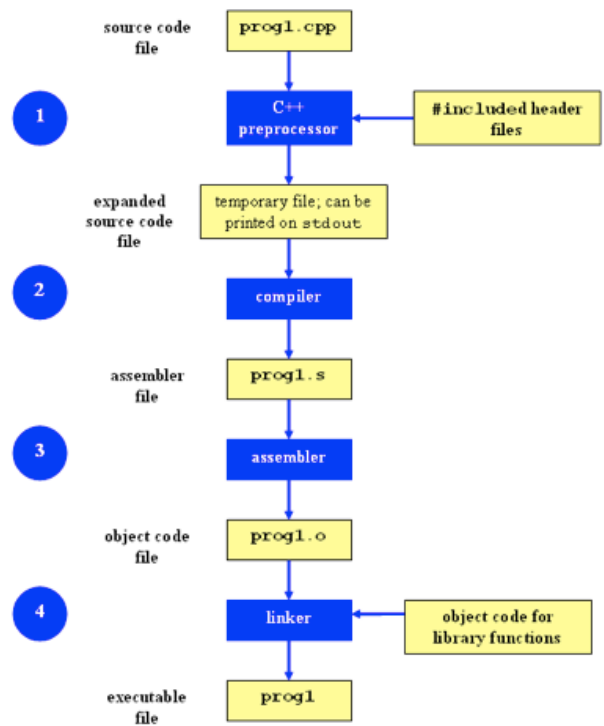

**Figure 2: C++ language Evaluation process**

When we compile your cpp file, we go through pre-processing by copying the content of the included header files in the source code file, generate macros, and replace #define constants with their values. Then, the compiler compiles the generated code into the assembly language of the platform. Right after, the compiler generates the assembler code into the object code for the specific platform [O-34]. This object code file is then linked together with the object code files of all libraries used to produce the executable file [O-35].

**Typing Strategy**:

C++ is a weak static typing strategy since it allows unsafe casts in exactly the same way C does [O-26]. C++ is a weakly typed language since it exposes pointers, as they are numeric values, which will be used to bypass the type system [O-17]. C++ is a static typing language since it supports objects with a predefined functionality. C++ static typing adds more constraints to the compiler. However, just like C, C++ cannot check everything while the program is running. C++ mixes weakly typed C code with static typed C++ to bypass all type checking. To demonstrate he C++ typing, the following example will be used [O-28]: We create a person object and make sure that the getName function always return a string of somebody's name.

```
class Person {

public string getName() {
        return "Stephane";
 }
};
```

Now, we can get our object as follows:

```
Person p;

printf("The name is %s\n", p.getName().c_str());
```

**Memory Management:**

In C++, it is possible to manage memory using different techniques to enhance program performance. There are four (4) techniques supported in C++ [O-19]. Static memory allocation is used to assign a static variable a value at compile time. The static keyword is used and the allocated storage is fixed throughout the program execution. Another technique use by C++ is automatic memory allocation. In this technique, an automatic variable is declared within its class name. When this variable is assigned a value, it is stored in the allocated stack [O-19]. When the declaration is executed, the constructor is called. Then, the destructor is called when the memory allocation needs to be freed. Dynamic memory allocation technique consists of manually manage memory by the software developper. The dynamic memory allocation is done through the keywords new and delete that are quite similar to C's malloc and free keywords [O-22], which are also supported in C++. One last technique used by C++ to manage its memory is garbage collection [O-22]. Bohem Garbage collection, known as Bohem GC, is used in C++. It allows finalization code to get invoked when an object is collected. Thanks to this garbage collector, developer could find memory leaks and double allocations [O-22].

### 1.2.5 Available Compilers/IDEs

C++ has several Integrated Development environments that come either as a built-in compiler or a separate compiler module. Some of these Compilers/IDEs can be summarized as follows:

**Qt** is a cross-platform application framework widely used to develop C++ graphical user interfaces as well as command line and consoles [O-47]. It is originally developed by Nokia and then maintained by Digia [O-47]. Qt uses C++ language combined with code generator (Meta Object Compiler) to enrich the language. It runs on several desktop and mobile platforms. The latest version Q5 was released on 19th December 2012 [O-48]. Using Qt, you can build strong C++ applications that can run in different platforms without code change [O-49]. It includes the Qt GUI module that provides classes to support system integration, event handling, OpenGL, 2D graphics, and imaging [O-50]. Moreover, Qt provides the Drag and Drop feature in which the user selects a GUI component and grabs it by dragging it into the form [O-51], which promotes easy and quick GUI development. Besides, Qt has non-GUI features like SQL database access, XML parsing, thread management, and network support. Qt is available for Windows, Mac OS, and Linux operating systems and can be downloaded from http://qt-project.org/downloads

**GTK+** is a multi-platform solution used to create graphical user interfaces [O-52]. It is written in C and designed to support a wide range of languages, such as C++. Along with Qt, it is one of the most popular toolkits. GTK+ was originally designed and used in GNU Image Manipulation Program (GIMP) to replace the Motif toolkit [O-53]. It originates from GTK and it is a free software graphics editor. One of the best characteristics of GTK+ is its syntax simplicity [54]. Everything is predefined in the form of libraries and we just need to use them. This allows GUI creation with a few lines of code, which promotes readability and makes bug fix and maintenance easier. GTK+ code is portable (Write once, work anywhere) and it stops C/C++ developers to turn to Java and gives them a real new alternative to learning a whole new multi-platform language. It is available for Windows, Mac OS, and Linux operating systems and can be downloaded from: http://www.gtk.org/download/

**WxWidgets** is a C++ library that lets developers create GUI applications for windows, Mac OS X, and Linux platforms and have their code compile and run on these platforms with minimal code change [55]. It uses the platform native API and therefore gives applications a truly native look and feel. Created by Julian Smart at the university of Edinburgh in 1992, WxWidgets is a free and open source distributed under wxWidgets Licence [O-56]. It supports a large number of compilers, such as Microsoft Visual Studio, Borland C++, Cygwin, and MinGW. WxWidgets library makes GUI programming easy and could shorten development time with powerful easy to use classes [O-57]. It has a WYSIWYG editor called wxFormBuilder that eases building user interfaces and design frames, panels, toolbars, and menu bars at edit time and then automatically generate the equivalent code in the file behind. It is available for Windows, Mac OS, and Linux operating systems and can be downloaded from: https://www.wxwidgets.org/downloads/

## 1.3 Visual Basic:

### 1.3.1 History & versions:

Visual Basic was first designed by two mathematician professors John Kemeny and Thomas kurtz at Dartmounth College in new hampshire in early 1960s[C-1]. The main idea to design this new language is to make it easy for college students to learn and use. The language first started as a part of the FORTRAN

11
A Comparative Studies of Programming Languages (Comparative Studies of Six Programming Language)

language but eventually they started creating a new language called BASIC[C-1].This basic language became so popular because it was easy to use[C-1]. In mid 1970s -1980s it became so popular that its started shipping which micro computers. This allowed small business owners, professionals to develop a customize software at affordable price [C-2].

General Electric used BASIC to design their largest time sharing system. In mid 1970s, Bill Gates and Paul Allen wrote one of the version of BASIC for the first microcomputers called the MITS Altair[C-1][C-2].The introduction of basic in Microsoft window's operating System made a new milestone for BASIC language. In 1991, VB1.0 was Released as it was the combination of code "Ruby" and "Embedded Basic". Ruby was developed by Alan Cooper [Who is known as The Father of visual Basic ][C-3][C-4].In 1987 allen cooper got an Idea while interviewing him as one client wants a wide range of shell solutions. So he decided to start creating a language name "Tripod" which is a set of tools where each user is able to customize his own application. During the development of tripord, Alan came up with the idea of drag and drop[C-5]. In 1988, Alan showed his tripord prototype to Bill Gates and bought it. After that a team of skilled programmer namely mark,Gray, mike and frank started to work on the stable release of a tripord which was now called as ruby.In 1990 after 18 months the langunage was ready to release and was tested by microsft's quality team [C-5].the original intention of shiping ruby was with windows 3.0 by microsoft did not shipped it with windows 3.0 they delay the shipping. Microsoft converted it from shell scripting to a visual programming by adding QuickBasic[C-5].In 1991 microsoft released VB1.0. This language brought evolution to high-level languages. VB language eventually has kept evolving and, nowadays, it is a main part of a .NET framework. Also, other languages like C# is highly nfluenced by Visual Basic.

**Versions:**

There are several versions of Visual basic as it was developed eventually.

**Visual Basic Learning Edition:**

It includes basic features of Visual Basic and Allows programmers to easily create powerful application for Windows.[C-14]

**Visual Basic Professional Edition:**

It allows professionals to develop solutions using the full featured set of tools. It includes all features of Learning Edition and it includes ActiveX Controls, Internet information Server application designer and Dynamic HTML Page designer.[C-14]

**Visual Basic Enterprise Edition:**

It allows creating robust distributed applications. It has all professional features. Also, it has support for back end tools such as SQL Server, Microsoft Transaction Server, Internet Information Server, Visual SourceSafe.[C-14]

Furthermore Visual basic had eventually evolved and in 1998 Visual basic 6 got released. VB6 was the stable version of visual basic, which became more popular. After that, Microsoft started on creating a .net framework and with the first version of .net framework they released new version of visual basic which is known as visual Basic. Net. By the time, VB has known enhancements and we currently have visual basic 2013 as the current version in .net framework 4.5

### 1.3.2 Paradigms

Visual Basic mainly used two programming paradigms Object Based and Object Oriented.Initially, Visual Basic was Object based programming language. That means that it has notion of objects, classes, Data Abstraction, Strong type checking but it does not have support for Inheritance, subtyping and polymorphism[C-6].Visual Basic 6 was using object based programming. As the language evolved, it transforms into object oriented language [C-1]. Visual Basic was evolved throughout the decade it grew as structure programming language in 1970s to make program easy to understand, develop and maintain[C-7]

Programming motivates the use of modularization where a developer can divide the large program into smaller parts. These creates functions that can be reused several times within the project[C-7]. Visual basic was object based and very popular at that time. Eventually, object oriented concepts became more popular and Visual basic needed to get updated. As a result, Microsoft released a new object oriented version of Visual basic which is known as a Visual Basic .NET. Visual basic evolved from COM programming to CLR compilation. Visual basic was first static and Strong type checking but as it evolve the type checking also evolve and Visual Basic.net it uses both strong and weak type checking. Type of various constructs like variable, functions expression can be check to reduce the bugs in programs and this can be done statically at compile time and dynamically at run time. Visual basic .net is capable of doing both type checking. Furthermore depending on application it works as strong type or weak type checking [C-8].

The most important Programming paradigm of Visual Basic is "Event-Driven".The basic concept of Event driven is that it partially allows programming logic that depends on user interaction to control program execution. The user triggers the event by clicking on either graphical control or keyboard keys. Each event has its own event handler that is related with a programming code. Event driven programming is preferred by most of users as the events are determined by end users, which allows them to interact with the application. It is widely used in graphical user interface design.[C-15]

### 1.3.3 Strengths and Weaknesses and application areas:

**Application Area**: Visual basic is now a part of a .NET framework. So, the developer can use VB.net to create a windows, Web and console applications. Developers can create large powerful high quality applications using VB.NET. Depending upon their needs, programmers can choose one of the options mentioned above to develop the application. VB.NET is widely used for in-house applications. It can also be used to create AJAX and COM components to use either online or on desktop application.

**Advantages** : Vb.net provides a rich GUI support and OOP concepts which makes it a competitive language to develop an interactive aplications. VB.net has Morden interface style which gives a good and simple look to developer to develop an application.It uses static type ckecking which analyize the source code and gives error like Immediate Syntax error, like missing characters, using wrong keywords. Ability to compile and run with in the IDE. It has Customization capabilities which allows user to highly customize IDE appearance[C-9].One of the interesting advantages of VB is its simplicity. It is an easy-to-learn language. VB is a component integration language which utilises Microsoft's component object model that allows any system part to be bolted onto program easily.[C-16]

**Disadvantages** : On the other hand, VB.net generates lots of automated code which makes it difficult for developer to understand. Visual basic is a proprietary programming language written by Microsoft, so Program written in visual basic cannot easily transfer to another language[C-10]. Visual Basic is a part of .net framework which has many other languages. So basically, VB starts to loose it competitiveness in front Microsoft languages, like C#, since they all share the same framework and they all could perform the same task[C-10].

### 1.3.4 Compilation and Execution:

Visual basic is not a part of the .net framework so IT will compile and execute in the same manner as other languages that run on .net framework like C#.

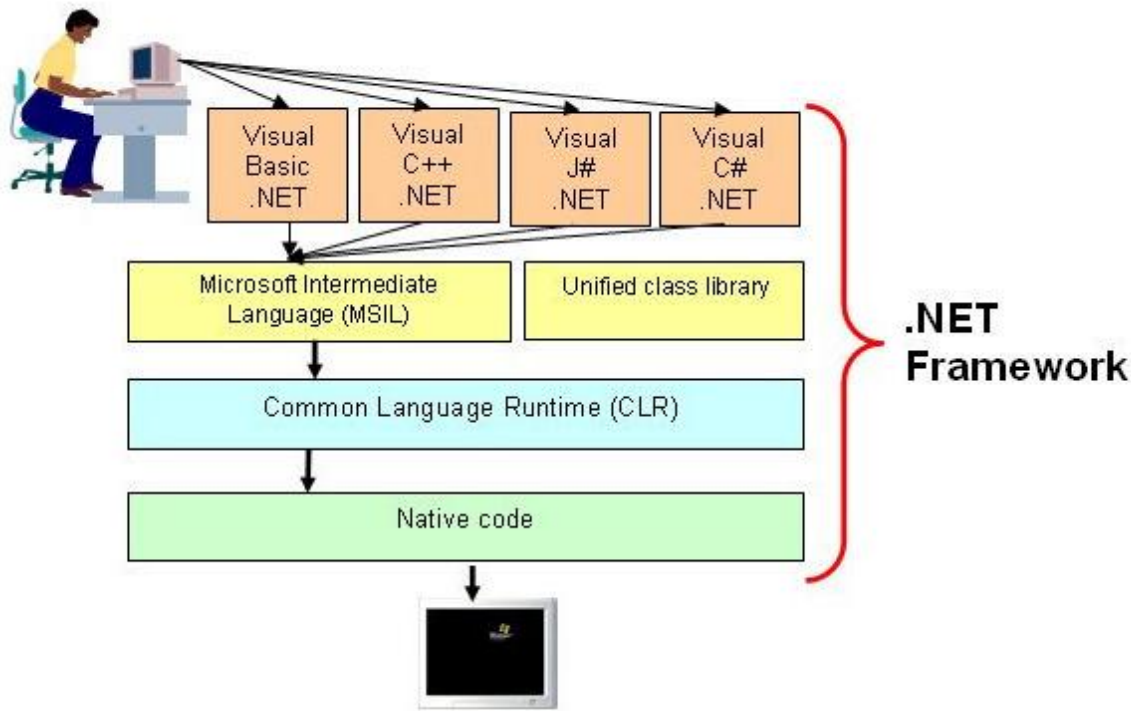

**Figure 3 : .NET Framework**

As we can see from the previous figure, languages using .net framework compile the code in the same way. For example the code written in visual basic language first gets converted into Microsoft intermediate language. This code get transferred to CLR which is common language runtime for .net framework and then converted into machine native code.[C-13]

**Memory management:**

Visual Basic has a Garbage collector similar to JAVA and C# which automatically manages the memory of the programs. Garbage collector runs automatically and checks for unused objects and releases its memory. When developer creates an object in C#, memory is automatically allocated. But if the object is no longer used, Garbage collector would release the memory. However, if a developer wants to manage memory manually, then he can use Global Alloc to allocate memory and GlobalFree to release the memory.

**Typing Strategy:**

Visual Basic is a Static type checking language. It verifies type safety of the program by analyzing the source code. This makes program faulty and allows many bugs to be caught early in the development phase. In general, developer gets all the errors at a run time so he can solve them early, which saves debugging time and make the program less error prone. Visual Basic is strongly typed language. It focus more on the type of a variable during a compile time.For example type of calling function object must be the similar type of the called function's object.

### 1.3.5 Available Compilers/IDEs

Since VB.net is a part of .net framework, which comes with the best IDE developed by Microsoft so far: Visual Studio. Visual Studio has the monopoly as a best IDE for VB.net. It provides rich GUI interface ,custom Controls, built-in Compiler, Integrated with IIS server so user just has to press start button to run the application and do not have to worry about other details. Visual Studio comes with different versions with different commercial solutions and it is free. For Example Visual studio 2013 ultimate, Visual Studio 2013 professional, visual Studio Premium. Visual Studio is not a open source product. you can download the IDE from this [http://www.visualstudio.com/downloads/download-visual-studio-vs]  link which can run on operation systems windows7/8/8.1/windows server 2008/2012. We are going to use MySql as database so we have to use mysql connector for .net to perform the database connectivity. For GUI, we used inbuilt components. Besides, there is another IDE called #develop(sharpDevelop). It is a free and open source development environment for .net platform. SharpDevelop has similar features as Visual studio. It has Syntax highlighting, auto completion menu, and toolbar features. It uses GUI technologies like ,windows form, Window presentation Foundation and Entity Framework[C-11][C-12]. It has many different versions like sharpdevelop 5, 4.4, 3.2 which can be download from this

[http://www.icsharpcode.net/OpenSource/SD/Download/] website. This IDE runs on Window7/8/8.1 operating System.

## 1.4 Java:

### 1.4.1 History and Versions:

Java is a predominantly static typed programming language (though it also supports dynamic typing for some OOPs concepts like polymorphism) developed by James Gosling, Mike Sheridan, and Patrick Naughton in the year June 1991 at Sun Microsystems [K-3]. Many of the constructs were influenced from its counterparts in C and C++ programming languages [K-3]. Though it was heavily influenced by C and C++, it was different in its own ways. It eliminated pointers as its designers thought that developers are using pointers the way it was not meant to be. Also, it dropped the headache of managing memory as it brought in automatic memory management with the introduction of Garbage Collection [K-1]. Another feature that made Java to stand out of its peers was that it favoured WORA philosophy (i.e. Write Once Run Anywhere) with the introduction of Java Virtual Machine [K-2] [K-3]. As a result, all a system needs to have is a JVM to run a java code. When it was introduced to the public, it came with a huge API that included Threading, Networking, GUI API's etc [K-7], which was so popular amongst the developer community which resulted in java being propelled to the #2 spot in sheer number of developers [K-6].

**Versions:**

**There are many versions of Java and they are as follows:**

**Java SE (Java Standard Edition):** This edition of Java includes only the core programming language which includes the JVM and the standard library. The latest stable version of Java SE is 8.

**Java EE (Java Enterprise Edition):** This edition of Java includes Java SE and also other features necessary for web development, database connectivity etc. The stable latest version of Java EE is 7.

**Java ME (Java Micro Edition):** This edition of Java is for mobile devices. This version of Java is optimized to run on mobile devices than Java SE. The stable latest version of Java ME is 3.4.

### 1.4.2 Paradigms:

Java is an excellent OOPs language. It offers encapsulation through access modifiers such as private, public, protected etc. which vastly improves the reliability and security aspects of the code. But when compared with other programming languages such as C++ and Python, Java doesn't support multiple-inheritance of classes though it supports multiple inheritance through Interfaces. Java also has all the characteristics mandated by the Structural and Imperative paradigms. Nevertheless, the strange thing about Java is that, though it doesn't support 'goto' statement which is in principle with the characteristics of a Structural Programming language, 'goto' is still a keyword in java [K-6].

Java also supports Procedural Programming. But java doesn't let us have global variables or methods outside the class, which could be attributed to one of its design goal (i.e. it should be "simple, object-oriented and familiar"). Thus, Java is more of a OOPs language than anything else.

Apart from the above mentioned paradigms, Java also supports Reflective paradigm with API's available to access/create/modify/add class members [K-5]. Java is also one of the few programming languages which supports creation of a truly multitasking applications as it has a huge API that supports creation/management/communication between threads and processes. It also offers many data structures that support atomic access and it also provides many utilities required for thread/processes synchronization like (Locks, Mutex, Semaphore,Monitor) [K-6].

### 1.4.3 Compilation, Execution and Memory Management:

**Compilation and Execution:**

The java code is compiled to bytecode by the java compiler and it is executed on the JVM. As a result, the execution speed was tremendously slow when compared with other languages like C++ [K-3]. But with the introduction of JIT Compiler, the execution speed was greatly improved and it almost matched the execution speed of C++ in some scenarios[K-2] [K-4].

With JIT compilation, only the portion/ fragment of the code that is needed by the compiler is compiled to machine code [K-16]. This technique would greatly speed up the execution speed of the programming language. It can also perform some caching if a particular piece of code needs to be called more times. While translation, the JIT compiler can also apply some optimizations based on the machine on which the runtime is executing [K-16].

**Memory Management:**

Since Java runs on JVM it can also be called as a Managed Language as the JVM takes care of all the resources that is needed by the program [O-52]. All the allocation and deallocation of memory is handled by the JVM through the Garbage Collector [O-53]. When a new object is created, the GC allocates the necessary memory, and once the object goes out of its scope, the GC doesn't release memory immediately but instead it becomes eligible for Garbage Collection [K-6], which would eventually release the memory.

**Typing Strategies:**

Java is a strongly static typed language but it also supports dynamic typing. Java supports dynamic typing thanks to the fact it supports polymorphism and reflection programming paradigms [K-15]. The advantages of Java being a static typed programming language, is that it allows programmers to find out the errors at the compile time[K-15]. Since Java is a strongly static type language, optimizations can be performed at the compile time as the types of objects would be known before hand, which would result in faster execution [K-15]. Another advantage of having Java as a static language is that a debugger would be able to identify what a particular piece of code would do, just by seeing the types of the input parameters and its return types [K-15]. One of the major disadvantage in having a strict static typed language is that we would end up in having so much "boiler-plate" code that doesn't contribute anything significant to the logic of the code.

A Comparative Studies of Programming Languages (Comparative Studies of Six Programming Language)

### 1.4.4 Strengths, Weaknesses and Application Areas:

Java is a good OOPs programming language, as it contains most of the OOPs concepts such as inheritance, polymorphism, encapsulation and it had also got rid off certain complicated concepts such as Multiple Inheritance due to reproduced issues such as the Deadly Diamond Death [K-14]. And hence the outcome of such decisions led to a simple and easy to use OOPs language.

Since Java is a static typed language, most of the coding errors are detected during the compile time because of which the unit testing time is greatly reduced and the user needs to test only the business logic [K-15]. Java also gives us the freedom in choosing from the vast library of quality frameworks available for any task that needs to be implemented using java.

Because of all these strengths, Java's usage is high in developing applications (predominantly web based applications) for enterprises. Java is also predominantly used in Education field to teach programming in colleges and universities as Java is an excellent OOP's language.

The most important weakness is that is that since it is not a natively executed language, its performance is a bit less when compared with the other natively executed languages such as C, C++. The memory footprint of Java is also a bit huge and hence because of the above weaknesses, Java is not suitable for real-time systems.

### 1.4.5 Available Compilers/IDEs

The compilers available for java is as shown below.

| Interpreters/Compilers | Development Status | Type | Platforms | Url |
|---|---|---|---|---|
| The official Java Compiler (javac) – This compiler comes with the programming language. | Active | Open Source | Windows, Mac OS, Linux | http://www.oracle.com/technetwork/java/javase/downloads/index.html |
| Jikes Compiler | No Active Development | Open Source | Windows, Mac OS, Linux | http://sourceforge.net/projects/jikes/ |
| Eclipse Compiler (ECJ) | Active | Open Source | Windows, Mac OS, Linux | http://www.eclipse.org/jdt/core/index.php |
| GNU Compiler | No Active Development | Open Source | Windows, Mac OS, Linux | http://gcc.gnu.org/java/ |

**Table 1 : compilers available for java**

There are many IDE's available for Java. And the most prominent ones are as shown below.

| IDE | Type | Platforms | Url |
|---|---|---|---|
| Netbeans IDE – A GUI based IDE used for developing applications for Java and many other programming languages through its plugins. | Open Source | Windows, Mac OS, Linux | https://netbeans.org/features/index.html |
| Eclipse IDE - A GUI based IDE used for developing applications for Java and many other programming languages through its plugins. | Open Source | Windows, Mac OS, Linux | http://www.eclipse.org/downloads/ |
| IntelliJ IDEA – A GUI based IDE used for developing applications for Java and other related programming languages like Groovy, Scala etc. | Commercial/Free Community Edition | Windows, Mac OS, Linux | http://www.jetbrains.com/idea/ |

**Table 2 : IDE's available for Java**

## 1.5 Python:

### 1.5.1 History and Versions:

Python is predominantly a dynamic typed programming language which was initiated by Guido van Rossum in the year 1989. The major design philosophy that was given more importance was the readability of the code and expressing an idea in fewer lines of code rather than the verbose way of expressing things as in C++ and Java [K-8][K-9]. The other design philosophy that was worth mentioning was that, there should be always a single way and a single obvious way to express a given task which is contradictory to other languages such as C++, Perl etc. [K-10].

Python compiles to an intermediary code and this in turn is interpreted by the Python Runtime Environment to the Native Machine Code. The initial versions of Python were heavily inspired from lisp (for functional programming constructs). Python had heavily borrowed the module system, exception model and also keyword arguments from Modula-3 language [K-10]. Pythons' developers strive not to entertain premature optimization, even though it might increase the performance by a few basis points [K-9].

During its design, the creators had conceptualized the language as being a very extensible language, and hence they had designed the language to have a small core library which was extended by a huge standard library [K-7]. Thus as a result, python is used as a scripting language as it can be easily embedded into any application, though it can be used to develop a full-fledged application.

The reference implementation of python is CPython. There are also other implementations like Jython, Iron Python which can use python syntax as well as can use any class of Java (Jython) or .Net class (Iron Python).

**Versions:**

Python has two versions 2.x version and 3.x version. The 3.x version is a backward incompatible release was released to fix many design issues which plagued the 2.x series. The latest in the 2.x series is 2.7.6 and the latest in 3.x series is 3.4.0.

### 1.5.2 Paradigms:

Python supports multi-paradigms such as: Object-Oriented, Imperative, Functional, Procedural, and Reflective. In Object-Oriented Paradigm, Python supports most of the OOPs concepts such as Inheritance (It also has support for Multiple Inheritance), Polymorphism but its lack of support for encapsulation is a blatant omission as Python doesn't have private, protected members: all class members are public [K-11]. Earlier Python 2.6 versions didn't support some OOP's concepts such as Abstraction through Interfaces and Abstract Classes [K-19]. It also supports Concurrent paradigm, but with Python we will not be able to make truly multitasking applications as the inbuilt threading API is limited by GIL (Global Interpreter Lock) and hence applications that use the threading API cannot run on multi-core parallelly [K-12].The only remedy is that, the user has to either use the multi-processing module which would fork processes or use Interpreters that haven't implemented GIL such as Jython or Iron Python [K-12].

### 1.5.3 Compilation, Execution and Memory Management:
**Compilation, Execution and Memory Management:**

Just like the other Managed Languages, Python compiles to an intermediary code and this in turn is interpreted by the Python Runtime Environment to the Native Machine Code. The reference implementation (i.e. CPython) doesn't come with a JIT compiler because of which the execution speed is slow compared to native programming languages [K-17]. We can use PyPy interpreter as it includes a JIT compiler rather than using the Python interpreter that comes by default with the python language, if speed of execution is one of the important factors [K-18]. The Python Runtime Environment also takes care of all the allocation and deallocation of memory through the Garbage Collector. When a new object is created, the GC allocates the necessary memory, and once the object goes out of its scope, the GC doesn't release memory immediately but instead it becomes eligible for Garbage Collection, which would eventually release the memory.

**Typing Strategies:**

Python is a strongly dynamic typed language. Python 3 also supports optional static typing [K-20]. There are a few advantages in using a dynamic typed language, the most prominent one would be that the code is more readable as there is less code (in other words has less boiler-plate code). But the main disadvantage in having python as a dynamic programming language is that there would be no way to guarantee that a particular piece of code would run successfully for all the different data-types scenarios simply because it had run successfully with one type. Basically, we don't have any means to find out an error in the code, till the code has started running.

### 1.5.4 Strengths and Weaknesses and Application Areas:

Python is predominantly used as a scripting language used in developing standalone applications that are being developed with Static-Typed languages, because of the flexibility it provides due to its dynamic typed nature. Python favours rapid application development, which qualifies it to be used for prototyping. To a certain extent, Python is also used in developing websites. Due to its dynamic typing and of the presence of a Virtual Machine, there is a considerable overhead which translates to way less performance when we compare with native programming languages [K-13]. And hence it is not suited for real-time/embedded applications.

### 1.5.5 Available Compilers/IDEs

Since Python language and its interpreter can be easily extended, there are several different packages of python available catering to different purposes. The most prominent ones are included below.

| Interpreters/Compilers | Type | Platforms | Url |
| --- | --- | --- | --- |
| CPython (Reference Implementation) | Open Source | Windows, Mac OS, Linux | https://www.python.org/download/ |
| PyPy – Python implementation | Open Source | Windows, Mac | http://pypy.org/download.html |
| having JIT compiler. | | OS, Linux | |
| Jython – Python implementation that runs on JVM capable of consuming JAVA code as well as python code. | Open Source | Windows, Mac OS, Linux. (Available as a jar which can be easily installed into JVM) | http://www.jython.org/downloads.html |
| Iron Python – Python implementation that runs on the .NET CLR, capable of consuming .NET code as well as python code. | Open Source | Windows, Mac OS, Linux (for Mac and Linux, Mono is needed) | http://ironpython.net/download/ |
| Skulpt – Python interpreter implemented in javascript, suitable for client side scripting in websites | Open Source | Windows, Mac OS, Linux | http://www.skulpt.org/static/proctest.html |

**Table 3 :Different packages of python**

There are also many ide's available for Python. A brief list is as shown below.

| IDE | Type | Platforms | Url |
|---|---|---|---|
| IDLE – A GUI based IDE that comes pre-installed with the reference implementation (CPython) | Open Source | Windows, Mac OS, Linux | https://www.python.org/download/ |
| PyCharm – A GUI based IDE by Jetbrains supporting code-completion, code-debugging, refactoring etc. | Commercial/Free Community Edition | Windows, Mac OS, Linux | http://www.jetbrains.com/pycharm/download/ |

A Comparative Studies of Programming Languages (Comparative Studies of Six Programming Language)

| | | | |
|---|---|---|---|
| PyDev – Eclipse Plugin supporting code-completion, code-templates, debugging etc. | Open Source | Windows, Mac OS, Linux | http://pydev.org/download.html |
| Aptana Studio – A GUI based IDE supporting other programming languages such as PHP, Ruby and JavaScript. This IDE supports C-completion, debugging and refactoring features. | Open Source | Windows, Mac OS, Linux | http://aptana.com/products/studio3/download |
| Komodo Edit - Powerful editor for Python, PHP, Perl, Ruby, Tcl, Javascript and other major web languages by ActiveState. | Commercial | Windows, Mac OS, Linux | http://www.activestate.com/komodo-edit/downloads |

**Table 4 : Different ide's available for Python**

## 1.6 C#:

### 1.6.1 History and Versions

C# programming language was introduced to the world by Microsoft at the same time as the .Net platform, and the different versions of C# were also introduced in parallel with the Microsoft .Net new versions. C# is a modern type-safe multi-paradigm programming language that became very popular and widely spread in the development field because of its simplicity, flexibility and productivity. The fact that C# benefits from several key features and powerful ideas found in different programming languages with different programming paradigms result in a programming language that can be used to develop applications with the clean syntax of Java, the simplicity of Visual Basic and the power and expressiveness of C and C++ programming languages [Z-1].

C# was influenced by three of the most widely used programming languages: C, C++ and Java. C and C++ are the direct predecessors of the C# programming language while there is a close relationship between it and Java programming language. Starting from C, the first structural programming language where power, elegance and expressiveness all came together, C# inherited its syntax, operators and a lot of keywords. Moving to C++, which was built as an enhancement to the C programming language to support the Object

Oriented Programming necessary to handle large programs, C# uses and further improves the object model defined by the C++ programming language. When talking about Java, which was motivated by the need of a platform-independent language to solve the portability problem emerged by the introduction of the Internet, C# adopted its elegant cost effective solution of using a portable intermediate language [Z-2]. Many programming languages were influenced by the C# programming language. Nemerle for example, which was designed to work on the .Net platform, has been influenced by C# in how it support Imperative and Object Oriented Programming. It also has a similar syntax to C# which reduces its learning curve [Z-13].

Together, Andres Hejlsberg and Scott Wiltamuth have written the C# language Specification [Z-3]. Anders Hejlsberg, the original author of Turbo Pascal and the chief architect of Delphi, joined Microsoft in 1996 and introduced the C# programming language to the world few years later as an initial release in 2000. In addition to his role as the chief designer of the new language, he was also involved in the development of the .Net platform [Z-4]. At the same time that Microsoft announced its first release of the C# programming language, it also proposed to the ECMA Technical Committee to standardize it [Z-5].

**Versions** [5]

- **C# 1.0 – Released January 2002**

This is the first version of the C# programming language where the basics of the language are defined. Important features supported by this version and kept by all the next versions include: the usage of the garbage collection to automatically manage the memory resources, operators overloading in a C++-like manner, the Attribute-Based Programming support and the ability to develop flexible applications without the need to deal with direct pointers anymore. This version described the C# formal syntax used to define classes, interfaces, structures, enumerations and delegates. The complete list of language specifications can be downloaded from:

http://download.microsoft.com/download/a/9/e/a9e229b9-fee5-4c3e-8476-917dee385062/CSharp%20Language%20Specification%20v1.0.doc

- **C# 1.2 – Released October 2003**

This version along the previous one are considered the trustful source of the grammar as well as the syntax of the C# programming language. This version of the language was released prior to the Microsoft Visual 2005. It contains description of new features that were added to the previous release but before the C# 2.0 that were introduced parallel with Microsoft Visual 2005. All aspects of the language are covered in detailed in the language specifications document. The C#1.2 language specifications can be downloaded from: http://download.microsoft.com/download/5/e/5/5e58be0a-b02b-41ac-a4a3-7a22286214ff/csharp%20language%20specification%20v1.2.doc

- **C# 2.0 – Released September 2005**

This version of the C# programming language introduced numerous features over the previous two releases. This version was released in parallel with the Microsoft Visual 2005. Among the huge set of features, some are notable such as the introduction of generics, iterators, partial and static classes, Nullable types, anonymous methods, covariance and contravariance in delegates and fixed-size buffers. The full specification of this version can be downloaded from:

http://download.microsoft.com/download/9/8/f/98fdf0c7-2bbd-40d3-9fd1-5a4159fa8044/csharp%202.0%20specification_sept_2005.doc

- **C# 3.0 – Released 2008**

By the release of this version, even more functionalities and features were added to the C# programming language. Supporting the LINQ strongly typed queries and the further support for the anonymous types, the introduction of the extension methods, partial method definition, object and the collection initializers, and the Lambda Expressions are all important features of C# 3.0 programming language. The complete language specifications is available to download from:

http://download.microsoft.com/download/3/8/8/388e7205-bc10-4226-b2a8-75351c669b09/CSharp%20Language%20Specification.doc

- **C# 4.0 – Released April 2010**

This version of C# programming language supports the dynamic type which provides the developer with new possibilities for simplifying the access to several COM APIs, dynamic APIs and the HTML Document Object Model. It also provides features such as the type equivalence and the covariance and contravariance support. The complete list of new features and specifications can be downloaded from:

http://www.microsoft.com/downloads/details.aspx?displaylang=en&FamilyID=dfbf523c-f98c-4804-afbd-459e846b268e

**Note:** The C# 4.0 specification is currently missing in action, so to speak - the previous link has been replaced by the 5.0 spec.

- **C# 5.0 – Released August 2012** [14]

C# most recent version 5.0 was released on August 15, 2012 with .NET Framework 4.5 and Visual Studio 2012. There are two main features in C# 5.0 - Async Programming and Caller Information [Z-15]. C# 5.0 Async feature introduces two keywords async and await which allows you to write asynchronous code more easily and intuitively like as synchronous code. Caller Information can help you in tracing, debugging and creating diagnose tools. It will help you to avoid duplicate codes which are generally invoked in many

A Comparative Studies of Programming Languages (Comparative Studies of Six Programming Language)

methods for same purpose, such as logging and tracing. The complete list of new features and specifications can be downloaded from:

http://download.microsoft.com/download/0/B/D/0BDA894F-2CCD-4C2C-B5A7-4EB1171962E5/CSharp%20Language%20Specification.docx

### 1.6.2 Paradigms

Besides being an Object Oriented programming language and the influenced of the C-family programming languages, C# adopted several functional programming languages features such as the anonymous types and the lambda expressions [Z-5]. Reasonable programming with C# in a functional programming style was revealed after the powerful features added to the language in its second version named C#2.0 such as generic classes and delegates and anonymous methods [Z-6]. By the release of the next version named C#3.0, the Lambda expressions introduced by this version provide further support to the functional programming in C#. What is more than that is the Language Integrated Query (LINQ) provided by the C# programming language version named C#3.0 to uniquely distinguish it from other programming languages in the field [Z-5]. Although C# was designed to primarily support imperative programming [Z-7], this programming language provides the developer with the ability to code in a declarative programming style by the use of the LINQ. Furthermore, the Reflection namespace provided by the C# programming language since the version 1.1 of its specifications and enhanced over time with each new version of the language, strongly support the reflective programming paradigm in C#. Building components using C# programming language is an easy task that helps greatly in managing a program's complexity. The properties, events and attributes constructs provided by the C# programming language makes it also a Component Oriented programming language [Z-3]. The most important component oriented feature supported by C# is its capability to work in a secure mixed-language environment [Z-2].

### 1.6.3 Strength/weaknesses and Application Areas

When C# was designed, one of its goals was to benefit from the strengths and avoid the weaknesses of other existing programming languages such as Java and C++ [Z-3]. C# is an efficient programming language that improves programmer's productivity by simplifying the development process and reducing the chances of having the most frequent programming errors. This language combine and balance the great about the two worlds of Java and C/C++ programming languages in such a way that allows developer to write programs in a simple elegant fashion as Java with the use of the expressiveness and flexibility of the C/C++ languages. The fact that it has similar syntax and grammar to the C++ and Java programming languages make it easier and faster to be learned and adopted. Since C# programming language was introduced after the introduction of many important Web standards such as the HTML, XML and SOAP, it was designed with the need to support Web standards and technologies in mind. In this field, a built-in support was included in C# programming language to allow for creating Web services capable of being invoked remotely regardless of the invoking application or its underlying platform. It is also worth mentioning, when talking about the supporting facilities of Web services in C#, that programmers can even use and control those Web services through the Web services framework the same way as they deal with C# objects. A flexible versioning support provided by the C# programming language eases the development and evolution processes of complex applications, increases their robustness and decreases their costs [Z-11]. The forced explicit interface implementations and the hiding of the inherited members are features provided by C# to support such flexible versioning [Z-3]. Furthermore, C# allows applying typed metadata to objects of any type such as classes and interfaces. This feature provides a great help in the process of monitoring and assuring correct mapping of the design into the implementation [Z-11]. Moreover, the component oriented nature of the C# eases significantly the process of managing and developing complex systems as a group of interacting components. Since C# is a type-safe programming language, it reduces common programming mistakes and enforces developer to adopt good programming behavior. Additionally, C# has a unified type system; meaning that every type is derived from one single base type. Thus the basic functionalities provided by this type are the minimum functionalities shared by any two types in C#. The garbage collector provided by C# automatically manages the memory resources at runtime. This feature also has its impact on reducing common programmers' mistakes when dealing with memory resources manually. What is different about C# is that the use of pointers is not completely eliminated rather than greatly reduced. C# provides developers with rich features that make the usage of pointers with no advantage for them except in performance critical applications [Z-3]. Since C# was designed specifically for the .Net framework, its supporting to the .Net features is more comprehensive and has a more suitable syntax than other languages [Z-12]. C# was designed to closely fit the .Net CLR and alias its type system and the mechanism used to handle exceptions and manage the memory [Z-3]. One of the major issues solved by C# is the cross-language interoperability, known also as the mixed-language programming. This interoperability between different pieces of code written in different programming languages is very important feature that is highly desirable, or even necessary, when implementing large distributed systems. Such a feature is also important in the case of implementing portable reusable software components that can be used by as many programming languages as possible and on numerous number of operating environments [Z-2].

Unfortunately, C# is still unsuitable to be used when developing time-critical applications or when the performance is required to be really high. Even though such applications are rare in the development field compared to other applications, this is a noticeable limitation of the C# programming language. Among the huge rich features and functionalities provided by C# it is still suffering from shortage in the key facilities required to develop high performance and time-critical applications such as specifying inline functions and destructors in such a way that they are guaranteed to run at particular points. The garbage collector also makes C# an inappropriate selection for such applications. It affects the performance of the application at runtime in an inconsistent manner at unpredictable points of time making it a hard task to develop applications that can produce consistent performance in real time [Z-12].

### 1.6.4 Evaluation Strategy
**Compiled and executed**

Very similar to Java programs, C# source code is compiled first into an intermediate language and then translated into machine code. When programs written by C# programming language are compiled, the result of the compilation process is an abstract binary format of the programs in an intermediate language, which follows the Common Language Infrastructure (CLI) specifications, called the Microsoft Intermediate Language (MSIL) which is similar to the Java independent bytecode. This MISL along with its resources are then stored on disk in an executable file with either .exe or .dll extension. This executable file, called an assembly, also provides description about its types, version, culture and security environments. The next step is done by the Common Language Runtime (CLR) which is a part of the .Net framework. In contrary to what is believed, the MSIL is not interrupted. Instead it is translated into machine code. When executing the C# programs, the executable assembly file is loaded into the CLR. This process may vary depending on the description information provided by the assembly file. When meeting all security requirements of the assembly is achieved, the MSIL is converted to the machine code that targets a specific machine by the CLR. To perform such conversion, the CLR performs a Just-In-Time (JIT) compilation process. In addition to the JIT compilation, several important services such as those related the garbage collection, resource managements as well as handling the exceptions are all provided by the .Net CLR [Z-8].

To conclude, two main phases are performed to compile and execute C# programs: the compile-time phase and the runtime phase. In the first phase, the C# compiler takes the C# source code files along with its resources and references and creates the assembly file. In the run-time phase, the .Net CLR takes this assembly file and converts it into the native machine code that can run on the target machine [Z-8]. The following diagram, taken from [Z-8], illustrates this process in a graphical representation.

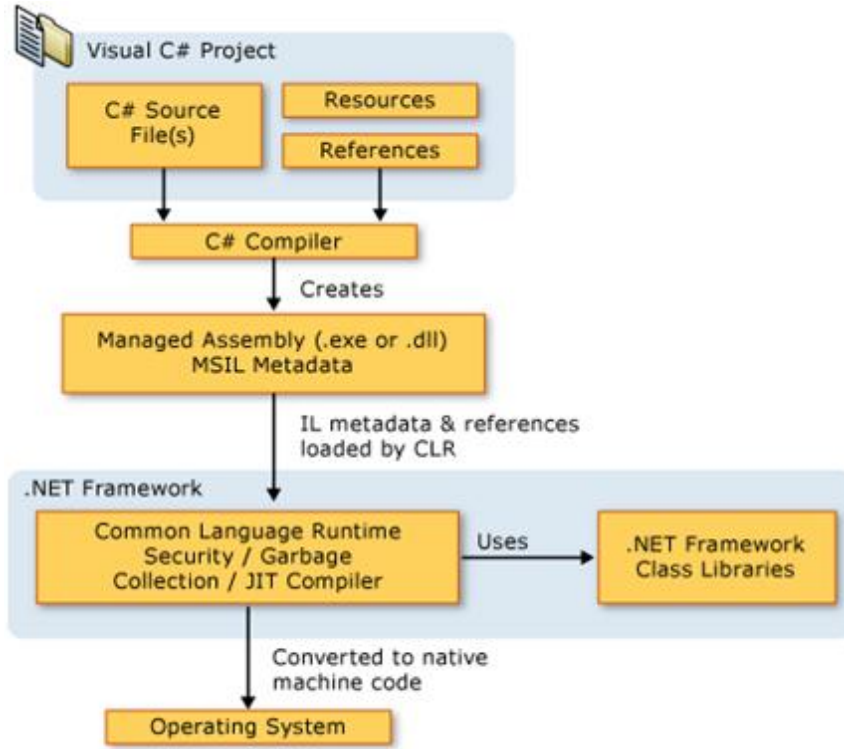

**Figure 4: compilation and execution in C# [Z-8]**

**Typing Strategies and Memory Management**

C# is a static strongly-typed programming language. A type must be specified explicitly to every variable and constant used in the code. Methods signatures also must explicitly specify the type to be return, and a type for every parameter. Types used within a C# code can be ready built-in types provided by the .Net framework class library as well as user-defined types. To insure a type-safe code, the C# compiler checks the operations performed on each variable against its type. Furthermore, the assembly file (the executable file results from the compilation step) also includes information about those types so it can be used by the Common Language Runtime for the purposes of managing memory resources at runtime [Z-9]. Although C# is a static typed language, the C# 4.0 and the Visual C# 2010 provides the new static type dynamic which provides C# with the ability to support dynamic types. Variables defined to be of this type are treated at compile time as if they can support any operation [Z-10].

### 1.6.5 Available Compilers/IDEs
**Integrated Development Environments and Compilers**

Several options and tools are available now a day to build C# applications. Some of those options are based on simple text editors along with the C# command line compiler while others are Graphical Integrated Development Environment that provide the developer with rich collection of features. By downloading and using the free .Net Framework 4.0 Software Development Kit, the developer can benefit from the variety of command line utilities and compilers and the .Net class libraries it provides. And building C#, or even any .Net, application can be done easily using the Notepad application along with the C# command line compiler called "csc.exe". Another IDE-free solution to build C# applications is the Notepad++ application. The Notpad++ is an open source tool that can be downloaded for free from: http://notepad-plus-plus.org/ . Noticeable features provided by the Notepad++ tool regarding developing C# applications include supporting C# keywords in terms of code coloring and auto-completion and syntax folding. The following sections will briefly describe the IDE solutions available to use in developing C# applications [Z-5].

**SharpDevelop**

It is a free open source development environment that is entirely written using the C# programming Language. It allows the developers to develop C#, VB.NET, IronPython, IronRuby, F# and Boo projects. Among the many features this IDE provides and are important to mention are: the designer utilities provided to visually design windows forms and databases, the object browsing and code definition utilities and the code-conversion utility that converts code from C# to VB and the other way around [Z-16]. Looking at the website of this IDE at: http://www.icsharpcode.net/OpenSource/SD/, four releases are available:

- SharpDevelop 1.1 that requires version 1.1 of the Microsoft .Net framework. An installed Microsoft .NET SDK 1.1 is also required. It can be downloaded from:
  http://www.icsharpcode.net/OpenSource/SD/Download/GetFile.aspx?What=Setup&Release=Fidalgo
- SharpDevelop 2.2 which requires version 2.0 of the Microsoft .Net framework. At least the .NET 2.0 runtime must be installed on the target machine. This release supports Windows XP SP2 and later, Windowas Vista and Windows Server 2003. It can be downloaded from:
  http://www.icsharpcode.net/OpenSource/SD/Download/GetFile.aspx?What=Setup&Release=Serralongue
- SharpDevelop 3.2 which requires at least .Net 3.5 runtime to be installed on the target machine. This release supports Windows XP SP2 and later, Windowas 7 and Windows Server 2008 R2. It can be downloaded from:
  http://www.icsharpcode.net/OpenSource/SD/Download/GetFile.aspx?What=Setup&Release=Montferrer

- SharpDevelop 4.2 which requires .Net 4 full runtime along with the .Net 3.5 SP1 runtime to be installed on the target machine. This release supports Windows XP SP2 and later, Windowas 7 and Windows Server 2008 R2. It can be downloaded from: http://www.icsharpcode.net/OpenSource/SD/Download/GetFile.aspx?What=Setup&Release=Mirador

**Visual C# Express**

This IDE is one of the free Visual Studio Express products that are available for free to allow developers simply create different .Net applications. Visual C# Express supports, as the name implied, .Net application developed using the C# Programming Language. Rich set of features and settings are provided in this IDE helping beginner as well as expert developers to create applications at multiple levels of complexity. In addition to all the features mentioned earlier in the SharpDevelop section, one important feature provided by the Visual C# Express is the support of the Windows Presentation Foundation applications [Z-17]. Available versions to download from the Microsoft website are:

- Visual C# 2005 Express which supports Windows 2000 SP4, Windows Server 2003 R2, Windows Vista, Windows XP and Windows XP SP2. It can be downloaded from: http://www.microsoft.com/download/en/details.aspx?id=804
- Visual C# 2008 Express which supports Windows XP SP2, Windows Vista, and Windows Server 2003. It can be downloaded from: http://www.microsoft.com/visualstudio/en-us/products/2008-editions/express
- Visual C# 2010 Express which supports 32 as well as 64 bits of several operating systems; Windows XP SP3, Windows Vista, Windows 7, Windows Server 2003 SP2 and R2 and Windows Server 2008 SP2 and R2. It can be downloaded from: http://www.microsoft.com/visualstudio/en-us/products/2010-editions/visual-csharp-express

**Visual Studio**

A premier commercial Microsoft's IDE that is available in different versions for different prices. The rich set of features provided by this IDE is far away from other free tools. In this IDE, visual XML designer and class designer are available to the developers in addition to the database and GUI designers. An integrated support for the code refactoring along with numerous set of other features are all available in this IDE as well [Z-18]. Several versions are available to purchase from Microsoft website [Z-19]:

- Visual Studio .Net 2003 which supports Windows 2000, Windows NT, Windows Server 2003 and Windows XP. It can be downloaded from: http://www.microsoft.com/download/en/confirmation.aspx?id=703
- Visual Studio 2005 which supports several operating system including Windows 200 SP4, Windows Server 2003, Windows Server 2003 R2, Windows XP, Windows XP SP2 and Windows Vista. It can be downloaded from: http://www.microsoft.com/download/en/confirmation.aspx?id=5553
- Visual Studio 2008 which support Windows Vista, Windows XP and Windows Server 2003. It can be downloaded from: http://www.microsoft.com/en-us/download/details.aspx?id=7873
- Visual Studio 2010 which support 32 as well as 64 bit and several operating systems including: Windows 7, Windows Server 2003 R2 and SP2, Windows Server 2008 R2 and SP2, Windows Vista SP2 and Windows XP SP3. Is can be downloaded from:
  http://www.microsoft.com/download/en/confirmation.aspx?id=23691
- Visual Studio 11 Beta which supports 32 as well as 64 bits along with Windows 7, Windows 8 Consumer Preview, Windows Server 2008 R2 and Windows Server 8 Beta. Different editions can be downloaded from: http://www.microsoft.com/visualstudio/11/en-us/downloads#vs

The free open source Xamarin's implementation of Microsoft's .Net framework, known as Mono, follows the ECMA C# and Common Language Infrastructure (CLI) standards. The idea behind designing such a software platform is to ease the development process of cross platform applications.

Mono consists of multiple components: the C# compiler, the Mono runtime, the base class library and the Mono class library. Different flavors of the C# compiler are available that support different sets of features of the C# programming language. The mcs C# compiler supports the full features of the C# 1.0 as well as C# 3.0 except generics and its dependant features. The gmcs compiler supports all the C# 3.0 features. The smcs compiler also supports the full features of the C# 3.0 but with the Silverlight APIs. The dmcs compiler supports the C# 4.0 features. The Mono runtime component, which is the Mono implementation of the ECMA standards for CLI, provides the developers with several useful services including a Just-In-Time compiler, an Ahead-of-Time compiler, a garbage collector and a library loader. The libraries provided by Mono provides the developer with a broad set of classes that provides functionalities similar to those provided by the Microsoft's .Net classes as well as additional functionalities offered by classes such as OpenGL, LDAP, Gtk+, POSIX and others.

About 74 releases divided into Legacy Branch and Current Branch of the Mono software platform is available to download at: http://mono-project.com/OldReleases.

The latest available *stable version is 2.10.8* which supports Windows, Mac OS X, Solaris, OpenSUSE and other. A *beta version 2.10.9* is also available to support Mac OS X is also available. Both releases can be downloaded from: http://www.go-mono.com/mono-downloads/download.html.

Xamarin also provides developers with two commercial solutions to use C# and the .Net framework to easy create applications for iPhone and iPad using MonoTouch for iOS; as well as for Android using Mono for Android.

**DotGNU Portable.NET**

The DotGNU Portable.NET is another free open source project available for C# .Net developers. Microsoft Windows, NetBSD, Solaris, Mac OS X are some of the supported operating systems by the DotGNU protable.NET project in addition to the prime target of the project which is Linux. The project includes a runtime engine called ilrun as well as a compiler called cscc. The runtime engine is implemented based on the ECMA standards of the Common Intermediate Language (CLI). This ilrun is used as an interpreter to programs in the CLI format. The compiler provided by this project supports C and C# programming languages. The front end of the compiler implements the ANSI C and the ECMA C# specification respectively. The compiler's back end implementation supports the ECMA C# specification definition of the CLI. The project also provides developers with a pure C# implementation of the controls of the System.Windows.Forms in such a way similar to Java Swing [Z-20]. The DotGNU Portable.NET project files can be downloaded from: http://www.gnu.org/projects/dotgnu/pnet-packages.html .

**Shared Source Common Language Infrastructure**

This Microsoft's project, also known as the Rotor project, is the implementation of the ECMA specifications for the C# programming language and the Common Language Infrastructure. This collection of source code files exceeds those printed ECMA standards, providing the opportunities to further understanding and investigation of the CLI implementation to both: researchers and developers. Included in this collection are the implementations of the CLI runtime, the Platform Adaption Layer (PAL), compilers that support the SSCLI for C# and development and build environment tools [Z-21].

Three versions available at the official website of Microsoft:

- Shared Source Common Language Infrastructure - Beta 1- which supports Windows XP operating system. It can be downloaded from:
  http://www.microsoft.com/download/en/confirmation.aspx?id=13221 .

- Shared Source Common Language Infrastructure 1.0 which supports Windows XP, Mac OS X 10.2 and FreeBSD operating system. It can be downloaded from:
  http://download.microsoft.com/download/.netframesdk/CLI3/1.0/WXP/EN-US/sscli_20021101.tgz

- Shared Source Common Language Infrastructure 2.0 which supports Windows XP SP2. It can be downloaded from: http://www.microsoft.com/download/en/confirmation.aspx?id=4917 .

## 2.1 Memory and Runtime Evaluations :

**Algorithms Used:**

We have used DFS, BFS and Kruskal's algorithms from Graph Theory to test the execution time and memory consumption in languages under study C++, Java, Python, PHP, VB and C#. We have used recursive algorithms for BFS and DFS as it would put considerable strain on the languages as the languages would have to store huge amounts of data including the content that resides in each of the local variables and it should also maintain the entire function call hierarchy. We have also used Kruskal's algorithm as it is a complex algorithm and it would give meaningful results with regards to performance as there are many looping and also frequent logical comparisons and memory swapping involved with this algorithm.

**DFS:**

**Pseudocode:**

*dfs_recursive(node):*

*push the node to stack and mark it visited*

*children = get all adjacent nodes to this node, from the graph*

*for child in children:*

*if child is not visited:*

*dfs_recursive (child)*

*pop the stack*

*return*

**Execution Time Comparisons (ms):**

| Nodes | Java | Python | C++ | VB | C# | PHP |
|---|---|---|---|---|---|---|
| 10000 | 49.09 | 26 | 23.5 | 53.9 | 53 | 57 |
| 20000 | 112.6 | 55.99 | 50.8 | 118.6 | 117.5 | 125.65 |
| 30000 | 130.84 | 87.99 | 70.4 | 145 | 144.9 | 165.32 |
| 50000 | 191.479 | 155 | 127 | 219.3 | 220.4 | 239.65 |
| 100000 | 403.65 | 312.32 | 259.5 | 431.34 | 412.27 | 474.8 |

Table 5 : Execution Time Comparisons in 6 Languages (DFS)

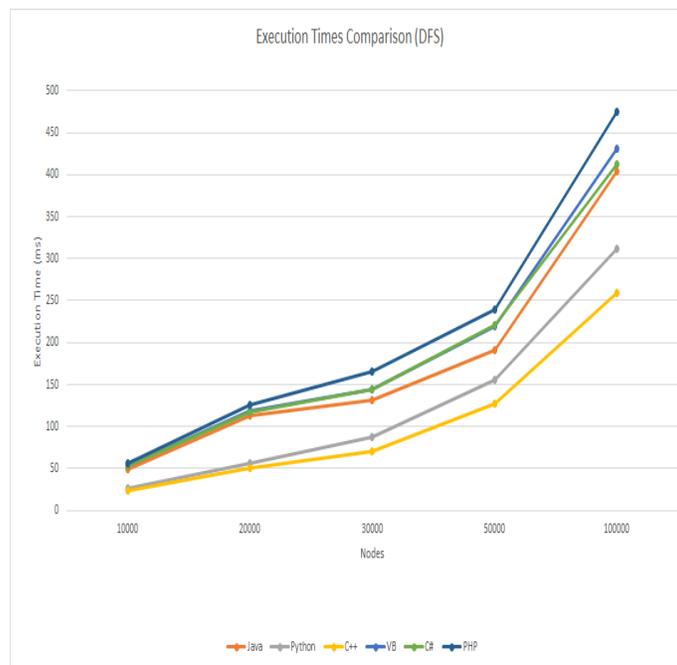

Figure 5 : Execution Time Comparisons in 6 Languages (DFS)

A Comparative Studies of Programming Languages (Comparative Studies of Six Programming Language)

**Memory Comparison (kb):**

| Nodes | Java (ms) | Python (ms) | C++ (ms) | VB (ms) | C# (ms) | PHP (ms) |
|---|---|---|---|---|---|---|
| 10000 | 21029.97 | 6799.36 | 1542.25 | 5342.39 | 5102 | 5723 |
| 20000 | 22688 | 8929.28 | 3149.27 | 7154.88 | 6898.2 | 7354.43 |
| 30000 | 29085.95 | 14622.72 | 6756.14 | 12652.73 | 12308 | 13698.79 |
| 50000 | 38642.79 | 18268.16 | 9563.38 | 15971.11 | 15350.64 | 17987.89 |
| 100000 | 50243.27 | 33754.84 | 16638.5 | 28724 | 28320 | 31024 |

Table 6 : Memory Comparisons in 6 Languages (DFS)

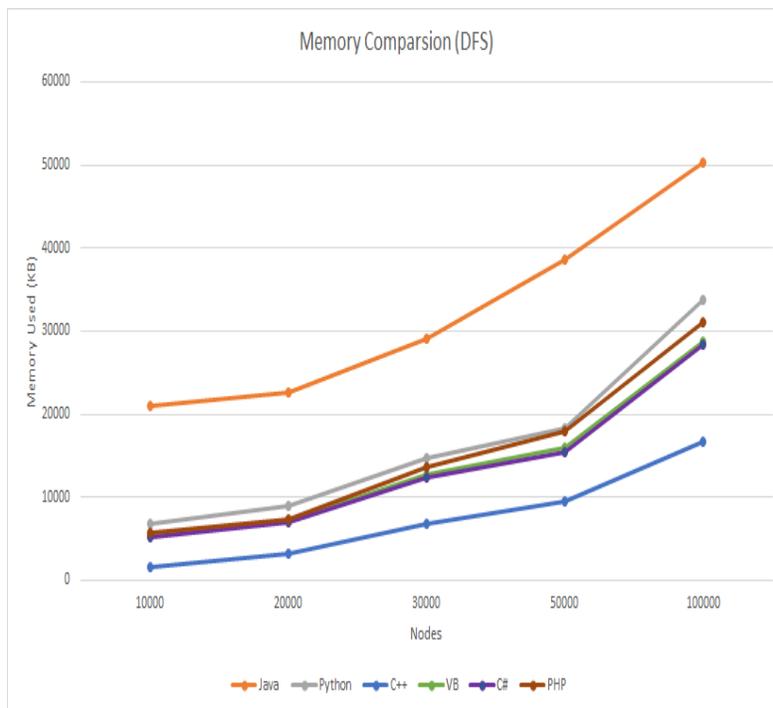

Figure 6 : Memory Comparisons in 6 Languages (DFS)

A Comparative Studies of Programming Languages (Comparative Studies of Six Programming Language)

**BFS:**

Pseudocode:

```
bfs_recursive (node):
    childNodes = get all adjacent nodes to this node, from the graph
    if node not present in visited:
        mark the node visited
    if childNodes not empty:
        for child in childNodes:
            If child not present in visited:
                add the child into queue
    if queue is not empty:
        bfs_recursive(dequeue the queue)
```

**Execution Time Comparisons (ms) :**

| Nodes  | Java    | Python | C++   | VB    | C#    | PHP   |
|--------|---------|--------|-------|-------|-------|-------|
| 10000  | 97      | 111    | 53.2  | 97.6  | 98.2  | 105   |
| 20000  | 125.67  | 219    | 79    | 134.2 | 134   | 184.5 |
| 30000  | 182.364 | 338.9  | 110.3 | 196.1 | 198.6 | 278.9 |
| 50000  | 253.19  | 680    | 169.7 | 276   | 277.4 | 535.2 |
| 100000 | 487.5   | 1143   | 344   | 579   | 535   | 871   |

**Table 7 : Execution Time Comparisons in 6 Languages (BFS)**

A Comparative Studies of Programming Languages (Comparative Studies of Six Programming Language)

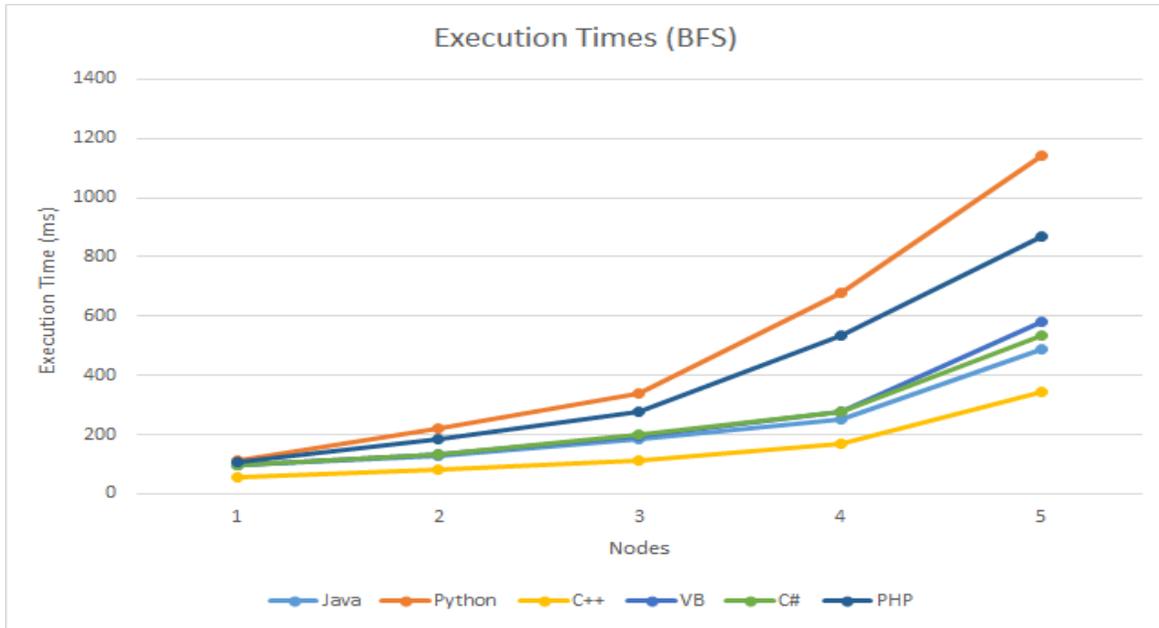

**Figure 7 : Execution Time Comparisons in 6 Languages (BFS)**

**Memory Comparisons (kb):**

| Nodes | Java | Python | C++ | VB | C# | PHP |
|---|---|---|---|---|---|---|
| 10000 | 9397.24 | 6830 | 2125.6 | 4365.9 | 4230 | 5698 |
| 20000 | 15565.35 | 8448 | 3674.8 | 6730.2 | 6524.9 | 7568 |
| 30000 | 29085.95 | 14436 | 7256.2 | 10245 | 10037.2 | 12984 |
| 50000 | 42479.34 | 18383 | 9867 | 14263.5 | 14536.9 | 17104 |
| 100000 | 55239 | 37245 | 17923 | 29143 | 28745 | 32634 |

**Table 8 : Memory Comparisons in 6 Languages (BFS)**

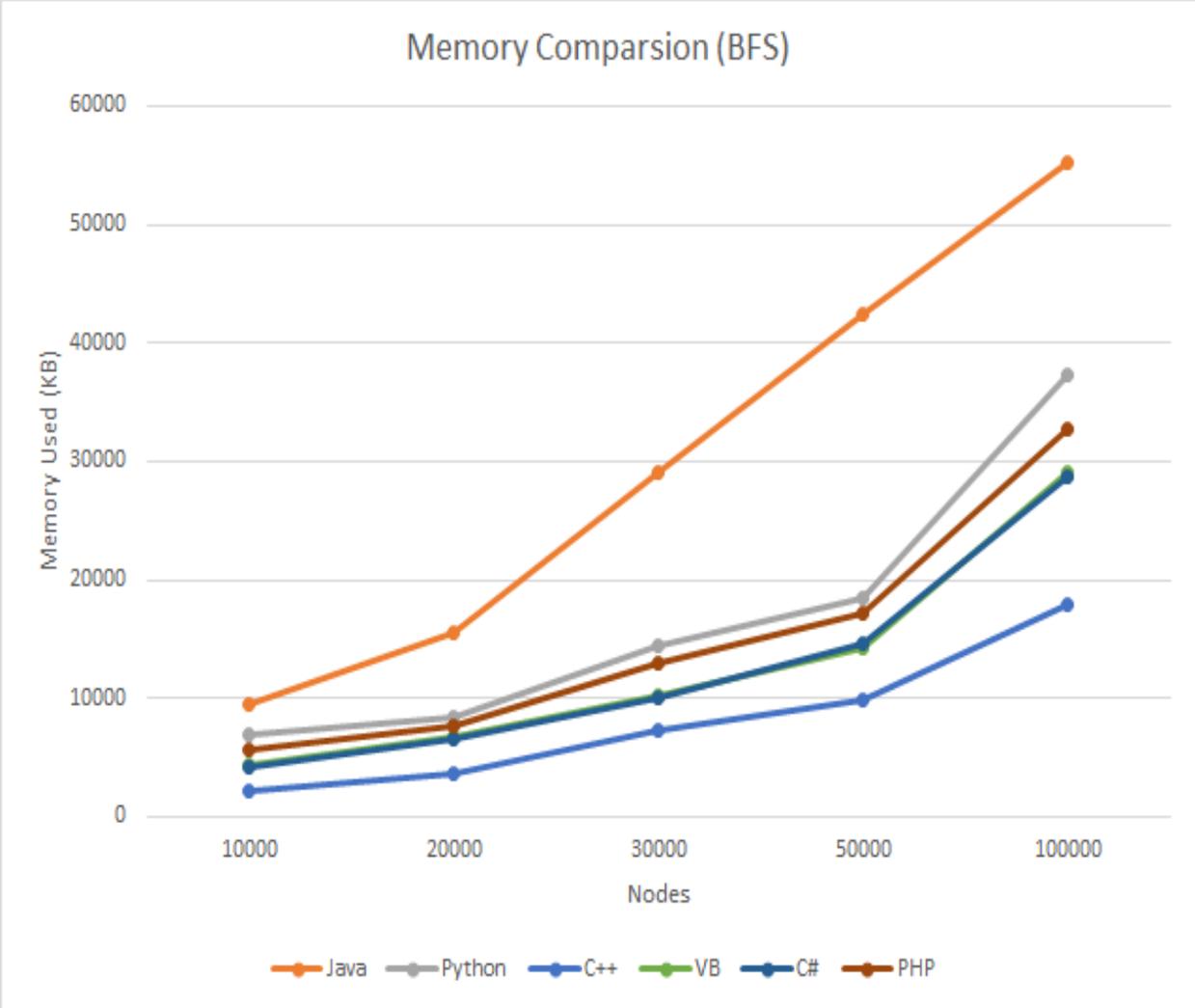

**Figure 8 : Memory Comparisons in 6 Languages (BFS)**

A Comparative Studies of Programming Languages (Comparative Studies of Six Programming Language)

**Kruskal's Algorithm:**

**Pseudocode:**

 1. Set A = {} and F = E, the set of all edges.
 2. Choose an edge e in F of minimum weight and whether adding e to A
    creates a cycle.
    (a) if "yes" remove e from F
    (b) if "no", move e from F to A
 3. If F={}, stop and output the minimal spanning tree (V,A).
    Otherwise go to step 2.

**Execution Time Comparisons (ms) :**

| Nodes | Java | Python | C++ | VB | C# | PHP |
|---|---|---|---|---|---|---|
| 10000 | 1386 | 1481 | 759 | 1409 | 1397 | 1423 |
| 20000 | 5289 | 6657 | 3645 | 5739 | 5512 | 5895 |
| 30000 | 11938 | 15690 | 7021 | 13198 | 13125 | 14260 |
| 50000 | 37324 | 49811 | 14124 | 39498 | 39652 | 46450 |
| 100000 | 60296 | 78249 | 29567 | 66124 | 65369 | 73451 |

**Table 9 : Execution Time Comparisons in 6 Languages (Kruskal's Algorithm)**

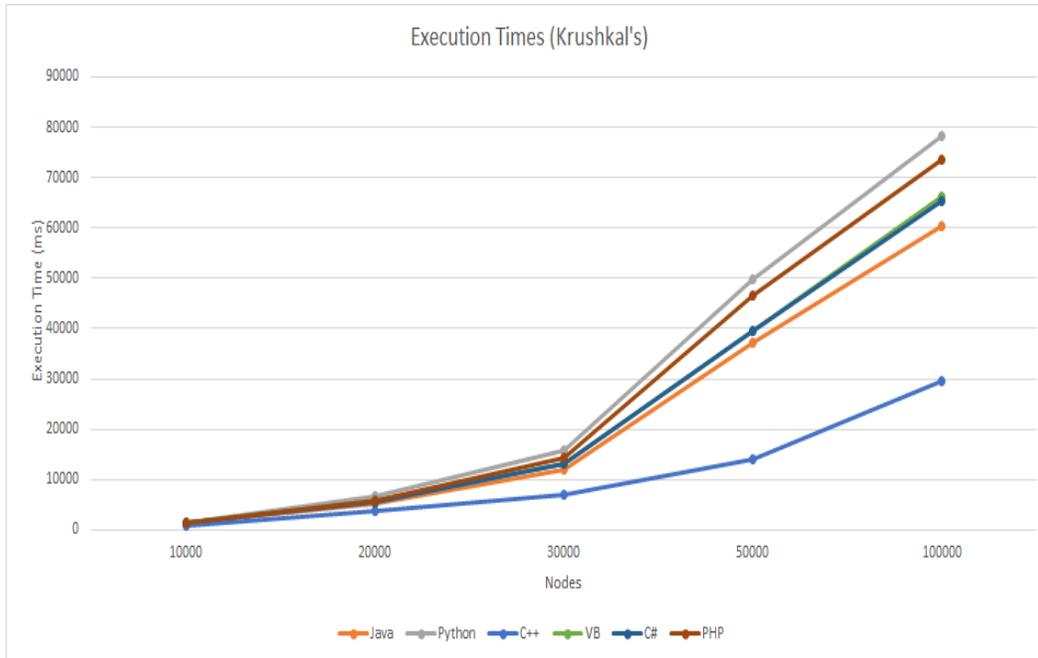

**Figure 9 : Execution Time Comparisons in 6 Languages (Kruskal's Algorithm)**

**Memory Comparisons (kb):**

| Nodes | Java | Python | C++ | VB | C# | PHP |
| --- | --- | --- | --- | --- | --- | --- |
| 10000 | 12650.89 | 6154.24 | 2983.5 | 4130 | 4356 | 5345.64 |
| 20000 | 19599.34 | 8642.56 | 4987 | 7469.29 | 7123.5 | 7989.25 |
| 30000 | 34010.26 | 10936.32 | 7689.25 | 8736.34 | 8412.79 | 9540.29 |
| 50000 | 92086.09 | 35473 | 19478 | 24327 | 23458 | 30478 |
| 100000 | 123415 | 73456 | 45236 | 55645 | 51321 | 67498 |

**Table 10 : Memory Comparisons in 6 Languages (Kruskal's Algorithm)**

A Comparative Studies of Programming Languages (Comparative Studies of Six Programming Language)

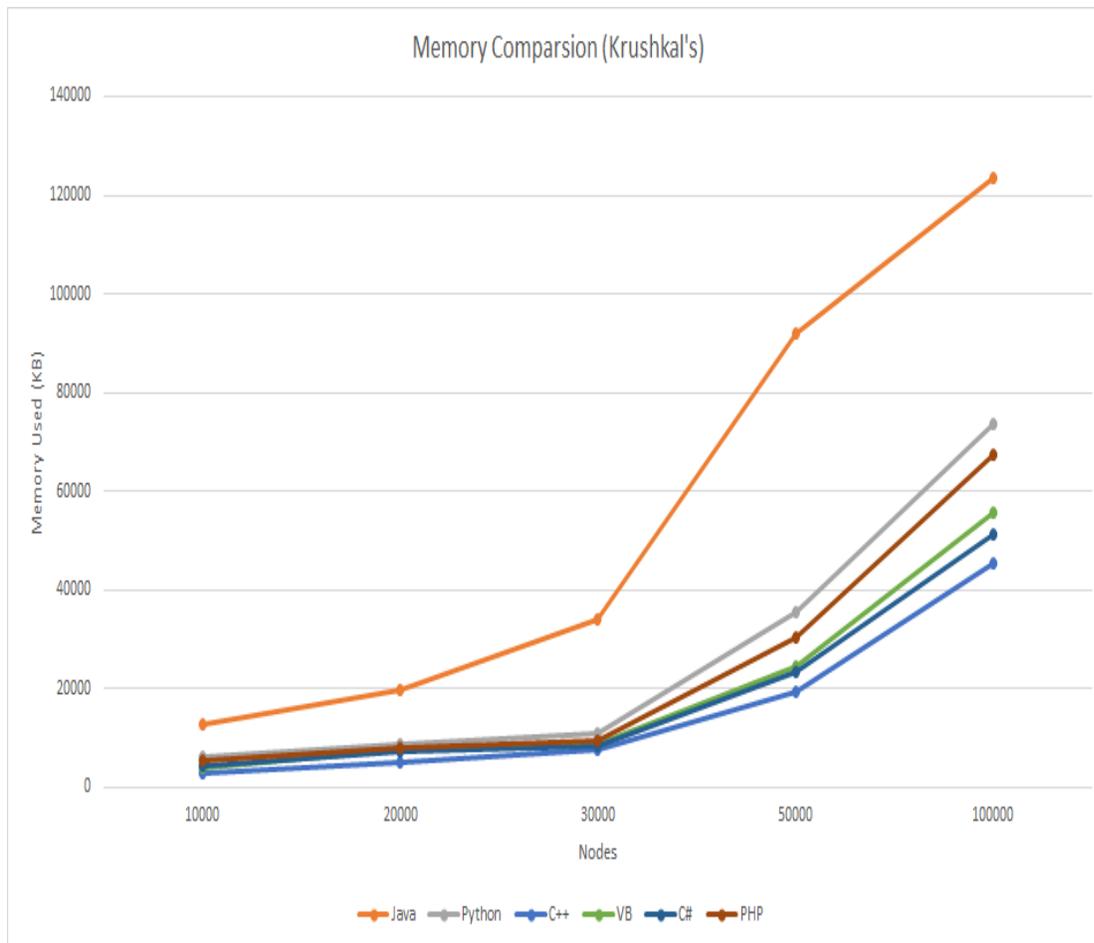

**Figure 10 : Memory Comparisons in 6 Languages (Kruskal's Algorithm)**

**Analysis:**

From the above graphs, it is evident that there are separate demarcations between intermediate level languages (C++), high level languages (Java, C#, VB) and dynamic typed high level languages (Python and Php). We notice that none of the above languages can match the speed and memory efficiency as C++. The reason why C++ has good efficiency is that it is weakly typed when compared to Java, C# and VB, meaning all the data are finally stored as Integer (ascii code) because of which the C++ runtime will not check whether a conversion from one datatype to the other is valid [O-13]. For instance, in Java, trying to convert a char to int will throw an exception at the runtime but C++ will not [K-6] . Since, C++ is not doing any type checking at runtime there could have been some gains in the execution speed. Another performance benefit for C++ is that it doesn't have a VM. This means that there will be significant improvement in speed of execution as the runtime has to run the machine code directly rather than converting an intermediary code to machine code and then run the machine code that was generated, as being done by the current generation VM [K-2]. The other high level languages that are used in the experiments have to take care of array bounding, illegal memory access and also memory allocation and deallocation which translates to slower execution than C++ [K-6][Z-2]. From the graphs, we can also see that the execution speeds of Java, C# and VB are almost related.

But Java has an edge when compared with C# and VB which could be attributed to Java having an optimized runtime [K-21]. C# and VB have almost the same execution times and memory consumption as both languages run on the same platform [K-22]. But C# has a slight edge when compared with VB as VB by default has checked arithmetic operations but whereas C# doesnot [K-22]. We can also interpret from the above graphs that dynamic typed high level languages such as Php and Python have less performance because of the overhead due to dynamic typing. A dynamic typed programming language cannot perform many optimizations as a strongly typed programming language would do, as there is no type information that is present at compile time and hence the language can do optimizations only when the program starts executing [K-15]. When it comes to memory consumption, java requires a huge memory footprint when compared with other languages because it has to load several classes to support various other services like (RMI, GC etc.) and it also has to load many classes from the vast java standard library, during the startup of a program [K-23]. The memory used is also high mainly because java has made many trade-off choices. As a consequence, the startup memory requirement and start-up time is higher when compared to other languages but the end result is that it can execute at near-native performance[K-23][K-17].

## 2.2 Database Connectivity (experiments):

For Database connectivity we have used "MYSQL" database as an external database which is used to connect with all our languages VB,C#,JAVA,Python,C++ and PHP. We have done reading and writing operations using MYSQL database to check out the database performance of each language. To do this we have decided to insert a predefined number of records with predefined memory (size of records). All the languages used the same database with same table schema which allowed us to compare the different languages with the same criteria. As we are using MYSQL database and not all the languages are feasible with MYSQL database directly. For example C# and VB need "mysql connector" connector to connect with the MYSQL database.For JAVA we used JDBC with MYSQL connector and for python we used MYSQLdb to connect with MYSQL database. For C++ we used mysql Connector/C to connect with database. Based on the specification mentioned above we implemented the code in all languages and we came out with the results which are mentioned below in the table .

A Comparative Studies of Programming Languages (Comparative Studies of Six Programming Language)

**Execution Time for Writing Operation with fixed Size of records:**

| Number of Records | C# | VB | JAVA | Python | C++ | PHP |
|---|---|---|---|---|---|---|
| 1000 | 1177 | 1259 | 1377 | 1437 | 920 | 1645 |
| 10000 | 12482 | 13059 | 12356 | 13987 | 11231 | 14675 |
| 50000 | 62987 | 66453 | 58810 | 70239 | 54789 | 76294 |
| 100000 | 125576 | 134980 | 112073 | 139217 | 113267 | 149721 |

**Table 11 : Execution Time for Writing Operation in 6 Languages (with fixed Size of records)**

**Graph Presentation:**

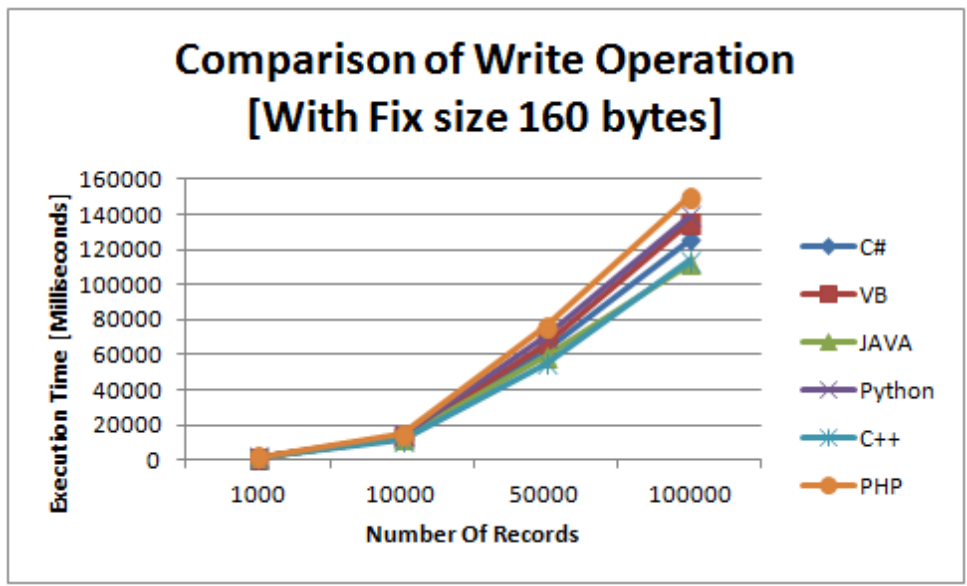

**Figure 11 : Execution Time for Writing Operation in 6 Languages (with fixed Size of records)**

A Comparative Studies of Programming Languages (Comparative Studies of Six Programming Language)

**Execution Time for Reading Operation with fixed Size of records:**

| Number of records | C# | VB | JAVA | Python | C++ | PHP |
|---|---|---|---|---|---|---|
| 1000 | 9 | 10 | 21 | 30 | 7 | 15 |
| 10000 | 127 | 138 | 229 | 250 | 120 | 187 |
| 50000 | 496 | 500 | 504 | 610 | 410 | 510 |
| 100000 | 845 | 950 | 948 | 1023 | 743 | 890 |

**Table 12 : Execution Time for Reading Operation in 6 Languages (with fixed Size of records)**

**Graph Presentation:**

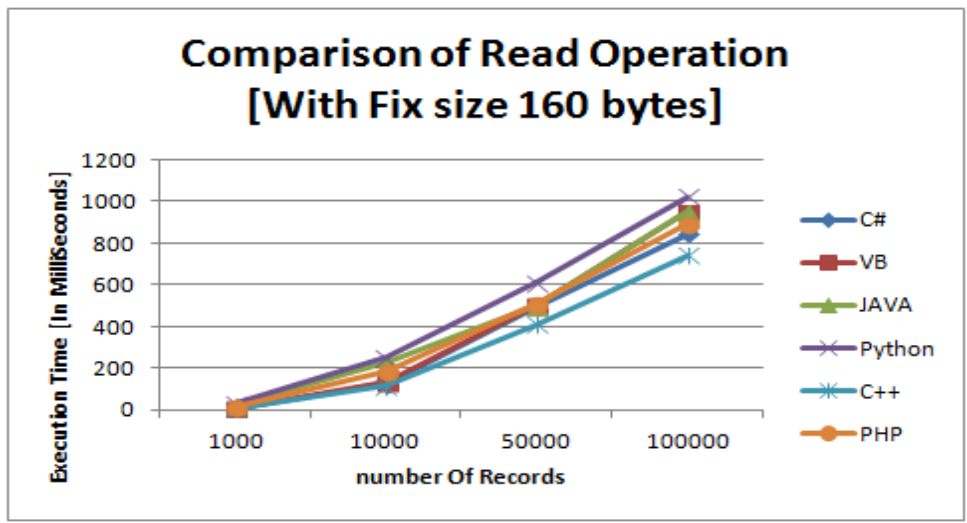

**Figure 12 : Execution Time for Reading Operation in 6 Languages (with fixed Size of records)**

A Comparative Studies of Programming Languages (Comparative Studies of Six Programming Language)

**Execution Time of write operation with fixed number of Records[10000 records]:**

| Size of Records | C# | VB | JAVA | Python | C++ | PHP |
|---|---|---|---|---|---|---|
| 4 | 10169 | 10523 | 12513 | 12714 | 9447 | 12945 |
| 40 | 11154 | 11965 | 12472 | 12910 | 10140 | 13102 |
| 80 | 10397 | 11587 | 13057 | 13479 | 10650 | 13879 |
| 160 | 11667 | 12098 | 13084 | 13921 | 11166 | 14220 |
| 255 | 12059 | 12743 | 13769 | 14121 | 12060 | 14978 |

**Table 13 : Execution Time for Writing Operation in 6 Languages (with fixed number of Records)**

**Graph presentation:**

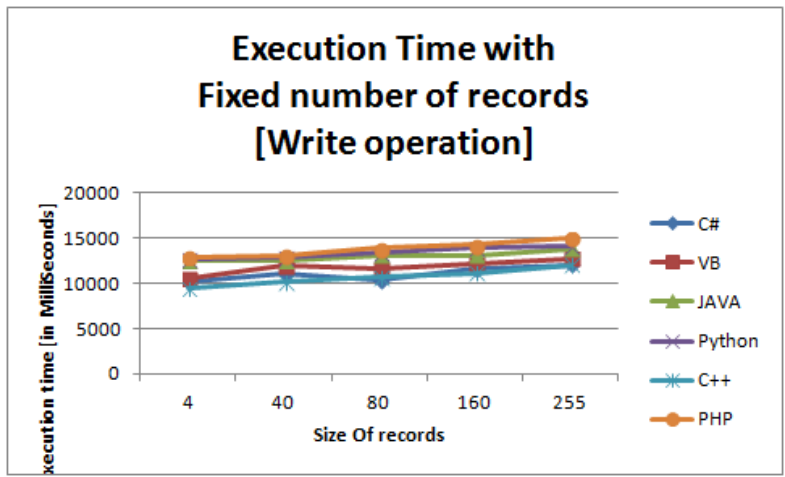

**Figure 13 : Execution Time for Writing Operation in 6 Languages (with fixed number of Records)**

**Execution Time of Read operation with fixed number of Records[10000 records]:**

| Size of Records | C# | VB | JAVA | Python | C++ | PHP |
|---|---|---|---|---|---|---|
| 4 | 31 | 36 | 100 | 121 | 27 | 73 |
| 40 | 31 | 39 | 151 | 179 | 29 | 129 |
| 80 | 42 | 50 | 188 | 211 | 38 | 210 |
| 160 | 63 | 78 | 410 | 489 | 59 | 379 |
| 255 | 156 | 183 | 559 | 584 | 103 | 510 |

Table 14 : Execution Time for Reading Operation in 6 Languages (with fixed number of Records)

**Graphs:**

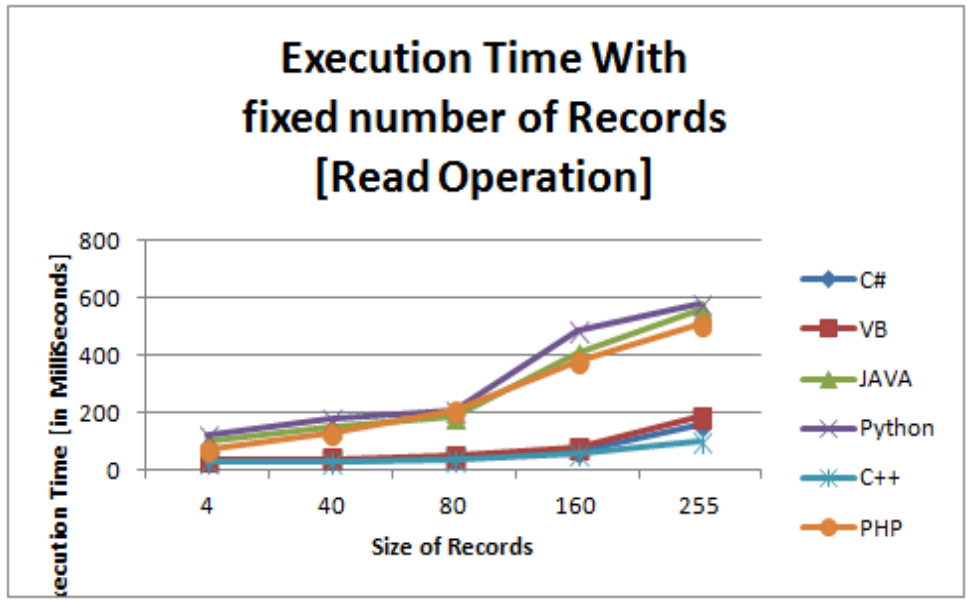

Figure 14 : Execution Time for Reading Operation in 6 Languages (with fixed number of Records)

A Comparative Studies of Programming Languages (Comparative Studies of Six Programming Language)

**Execution time for Write operation:**

| Number of Records | Size of Records[Bytes] | VB | C# | JAVA | Python | C++ | PHP |
|---|---|---|---|---|---|---|---|
| 1000(4) | 4 | 670 | 612 | 327 | 328 | 175 | 721 |
| 10000(40) | 40 | 7378 | 4148 | 3781 | 7147 | 2167 | 7500 |
| 50000(80) | 80 | 8269 | 5639 | 5134 | 7925 | 3892 | 8350 |
| 100000(160) | 160 | 12407 | 10731 | 10109 | 12510 | 9543 | 13890 |

**Table 15 : Execution Time for Writing Operation in 6 Languages (Increasing size and number of records)**

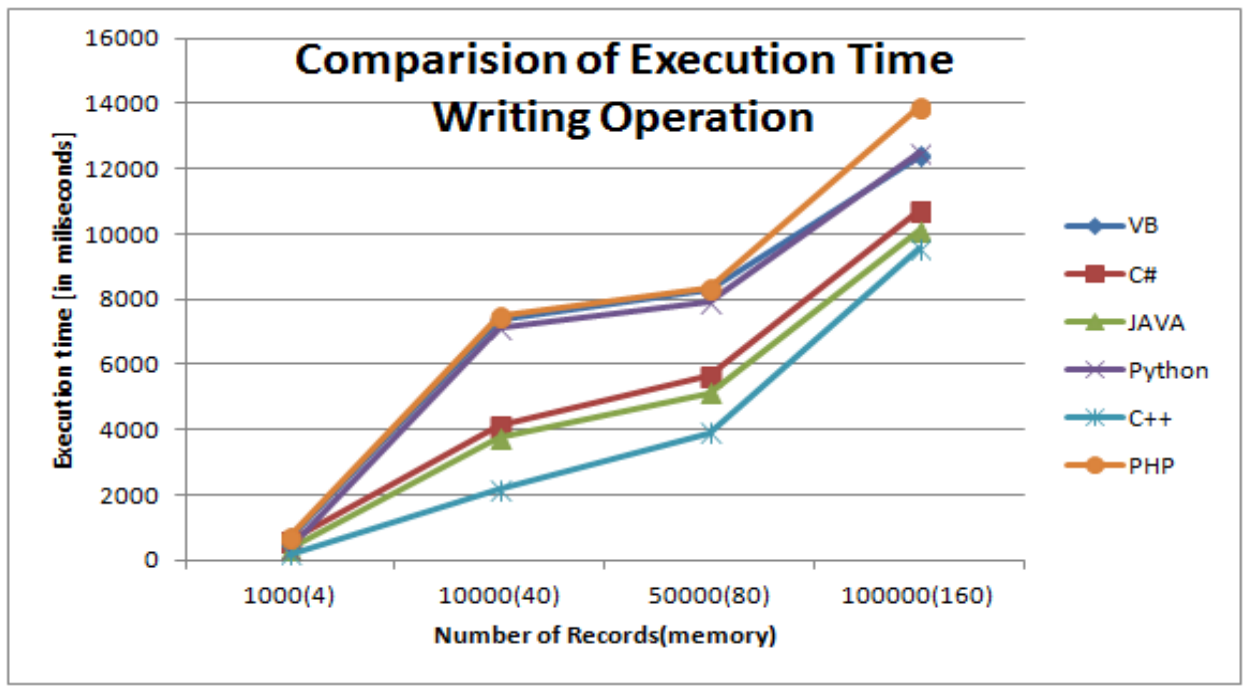

**Figure 15 : Execution Time for Writing Operation in 6 Languages (Increasing size and number of records)**

A Comparative Studies of Programming Languages (Comparative Studies of Six Programming Language)

**Execution Time for Read operation:**

| Number of Records | Size of Records[Bytes] | C# | VB | JAVA | Python | C++ | PHP |
|---|---|---|---|---|---|---|---|
| 1000(4) | 4 | 6 | 5 | 16 | 32 | 3 | 7 |
| 10000(40) | 40 | 38 | 44 | 47 | 50 | 24 | 40 |
| 50000(80) | 80 | 434 | 500 | 620 | 790 | 320 | 550 |
| 100000(160) | 160 | 2146 | 1436 | 2250 | 2439 | 1245 | 1987 |

Table 16 : Execution Time for Reading Operation in 6 Languages (Increasing size and number of records)

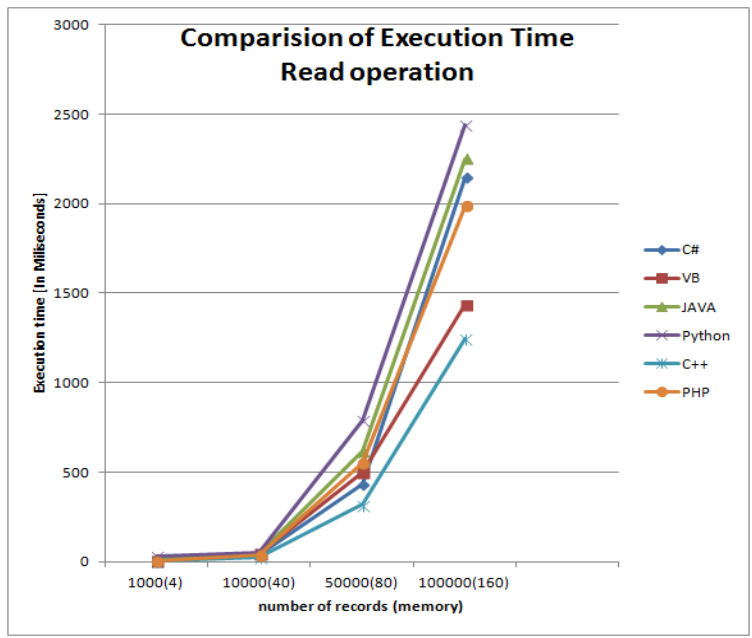

Figure 16 : Execution Time for Reading Operation in 6 Languages (Increasing size and number of records)

A Comparative Studies of Programming Languages (Comparative Studies of Six Programming Language)

**Interpretation of results**

**Writing Operation:**

We have created three Different Experiments with Write operation for each language and we came up with an interesting result. First we did experiment in which we kept the size of the records same and we just changed the number of records. For example we kept the size of records is 160 bytes and we take different records like 1000,10000,50000,100000 to insert to the database. For the table we can see that while inserting n 1000 records we got almost similar results for C++,C# and VB language and JAVA,Python,PHP have the same results but there is a but gap between these two groups.But as we increase the number of records from 1000 to 100000 we see a bit difference as All languages    VB and Python take a long time to execute than other languages. So we can conclude that as we increase the number of records to insert into database C++ works better than any other language, C#,VB,JAVA and PHP works almost similar and Python  takes longer to insert into database than any other languages mentioned above.

**Second Experiment:** In second experiment we kept number or records fixed which is 10000 records and we changed the size of records like 4,40,80,160,255 bytes and we tested this with all languages.From the table we can see that when we insert 1000 records with 4 bytes we see the significant difference in execution time between JAVA and C#. We can see that JAVA,Python,PHP take almost the same time which is longer than C#,VB and C++. And it remains same even though for the big size of records like 255 bytes. Java,Python,PHP take similar time which is longer than C#,VB and C++ execution time. C++ works faster than all languages mention above.

**Third Experiment :** We change size and number of records at the same time .For example 1000 records with 4 bytes, 10000 records with 40 bytes and so on. From the above table and graph for write operation we can assure that over all C++ performance is excellent when compared with the other languages under consideration.  From the table we can see that when we insert 1000 records with the size of 4 bytes C++ executes it in the fastest manner. While JAVA and Python is almost at take the same time to execute the same query and C# and VB takes almost same time. But PHP takes a longer time than all other languages. But as we increase the number of records and the size of records increases it leads to very different results of each language for execution. For example for 1000 records n size of 4 bytes VB and C# executes it in almost in same time but as We increase the size and the number of records to 40 bytes and 10000 respectively VB takes much longer time than C# and as we go on and increases the size and number of records there is a significant gap difference between the execution time of two languages.  The same things apply in comparison of JAVA and Python. One of the interesting fact is that C# and JAVA are give almost the similar execution Time for each test cases. On the other side PHP takes the maximum time in to execute a write operation among all the languages and As we increases the number of records with the size we can see the significant difference in the execution time of PHP language. Among the all languages C++ gives the best performance for the execution time to write  number of records to database.

**Reading operation:**

**Experiment one:** As we discussed in write operation we fixed size of records and we changed number of records for different tests. From the table of reading operation 1 we can see that when we retrieved 1000 records from C#,VB and C++ have almost a same execution time but what is interesting here is the result of java language as it takes a long time to read records from the database and As we increase the size of records execution time of java is getting longer. C++ takes minimum execution time to read records from the database where else JAVA and Python take maximum execution time to to read from Database.

**Experiment two:** As we mentioned in write operation for this experiment we kept number of records constant which is 10000 records and we changed size of records for different tests. for this experiment we also came to know that java is way slower than other languages like C#,VB,C++ to retrieve data from database.As we can see from the table of reading operation 2 we can see that when we retrieved 4 bytes data of 1000 records  JAVA and PHP take longest time to retrieve data from database, while C#,VB and C++ as a bit faster than other languages. But as we increased our size od records to 255 bytes we observed that C++ is faster than other languages, C# and VB are almost similar and bit slower than C++. Surprisingly JAVA takes more time than C# to retrieve data. JAVA and python has almost similar execution time. PHP takes a longer time from all other languages to retrieve data from database.

**Experiment Three:** For this experiment we changed size and number of records for different test cases. For first test case we used 1000 records with 4 bytes size from the table we can see that C#,C++,VB,PHP take almost similar execution time to read data from database while JAVA and python take a longer time to perform the same task. But as we increment the number of records with size of records we found that C++ is bit faster than all other languages. PHP also get better and take less time than C#,VB and java to read 1000000 records of 255 bytes. C# and VB are almost similar while Java and python take more time to read from database as compare to other languages.

**Analysis:**

Based on the graphs, tables and analysis of results mentioned above we can conclude that C++ is faster than other languages in both write and read operations to the database. As C++ does not have Virtual machine, like C# or JAVA, it directly compiles the code into the native code [O-12] [O-13]. As a result, the execution is faster. We found interesting result for JAVA language as writing operation in JAVA is almost similar to C# and VB but when it comes to reading operation JAVA is much slower than other languages. This is due to the fact that Java, by default, result sets are retrieved and stored in the memory which is not good for a query with a large result set. C# and VB are bit slower as the code is first converted in to Intermediate language then it goes to CLR and after that it convert to native code, this transformation of code leads to overhead to the performance of language and make it a bit slower than C++ [C-8] [Z-2] . PHP is an interpreted language because of this it is slower than C++,C#,VB,JAVA [O-3]. Python takes a longer time to do both the database operation. We used mysql connector/c to connect to mysql database, which is written in C language. So python has to convert from C Data types to Python data types which puts a overhead on the performance. Because of this reason python is slower than all other languages.

## 2.3 GUI Development (experiments):

User interface design and development tools help us in developing user interfaces. They allow end users to interact with software systems through visual indicators. Their main objective is to make the system simpler, interactive and easy to use.

In this section, we are discussing the similarities and differences between multiple GUI development facilities used to build graphical user interface of the languages under study: Qt (C++), Eclipse (php), Swing using netbeans (Java), TKinter (Python), and C# (Visual Studio).

In fact, there are several types of user interfaces. Each comes with different features and characteristics. In this comparative study, we will compare GUI development based on three (3) primary GUI types:

Command Line interface, Graphical user interface, and Web based user interface.

**Command line interface**:

Command line interface is the native way to do things. Natively, C++ programming language uses command line in order to perform its basic input/output operations. Command line interface is a domain specific language [Z-31][Z-32]that is used to describe the options C++ programs support. Compared to other languages, C++ command line interface has no external dependencies nor a run time library. Also, there is no command line interface menu framework for user navigation: It is up to the user to create one. Python has a command line interface called a cmdModule [Z-31][Z-32]. CmdModule makes it easier to make graphical user interfaces for Python and has a couple of advantages compared to other languages as C++ and C#. One of CmdModule advantages is that it gives your python programs a Shell interface. Also, you can easily drive a text oriented program with the popen command, which allows the the Python application to get compiled automatically . Also, unlike other languages. the Python cmdModule provides text completion feature that guides the interpreter to suggest word completion without the overhead of pressing the TAB key . On the other hand, C# language has a command line interface feature too. It includes basic functionalities as well as advanced characteristics. C# interface command line provides access to three (3) basic data streams: Standard input, Standard output, and Standard error. Unlike other programming languages, C# command line interface could be used for automated testing , which would help in reducing resources needed for automation implementation. Java affords many command line interfaces. The most common java CLI is the natural CLI . Natural CLI offers several features that are unique to java. One of these features is being a pure java implementation and making command line parsing easy using annotations . Also, it supports fully localisations unlike C++ and Python. PHP supports CLI SAPI whose main objective is to develop

PHP shell applications. This PHP command line interface allows you to run individual PHP scripts and ad-hocs and to perform command-line scripting in the server side . The command line interface is written in C and can also be used with standalone graphical applications.

**Graphical user interface**:

**Qt:**

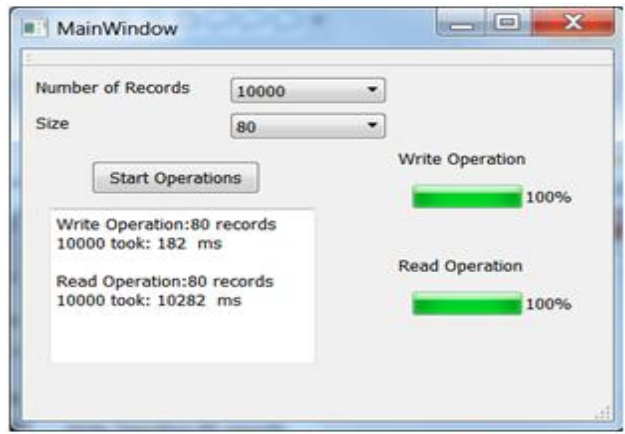

**Figure 17 : Main GUI's components using Qt.**

**PHP/HTML**

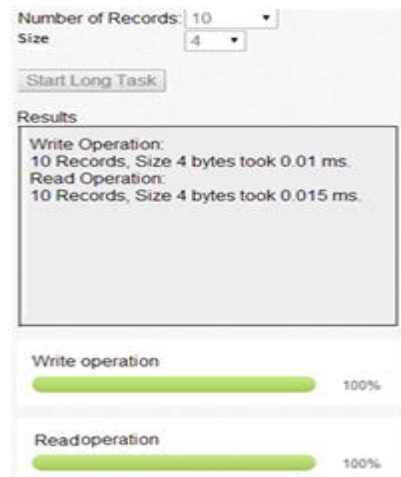

**Figure 18 : Main GUI's components using Eclipse + HTML (PHP).**

**C# development:**

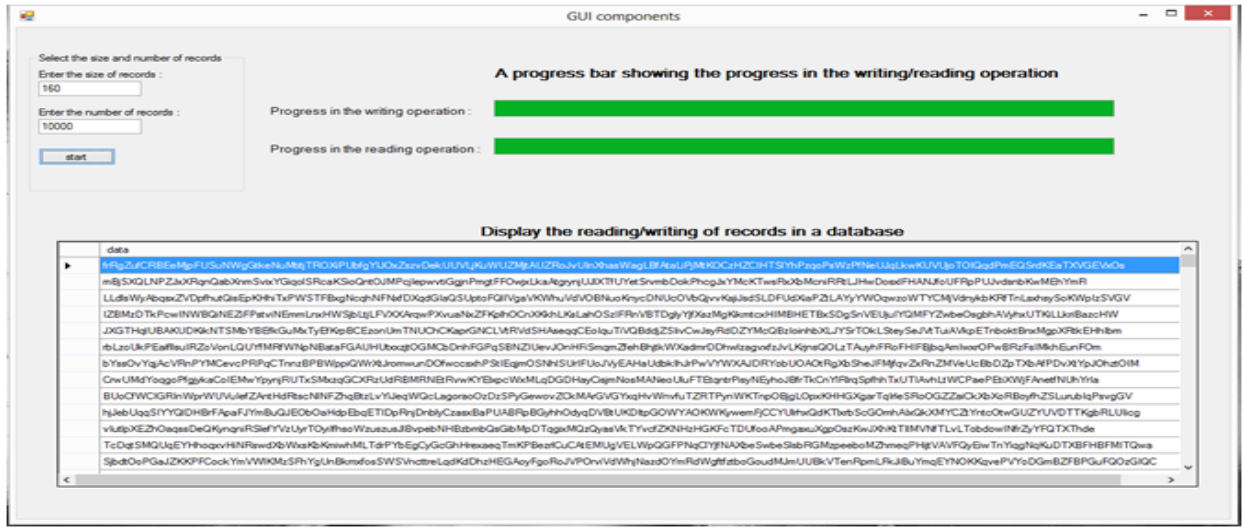

Figure 19 : Main GUI's components using Visual C#.

**Java development (Swing):**

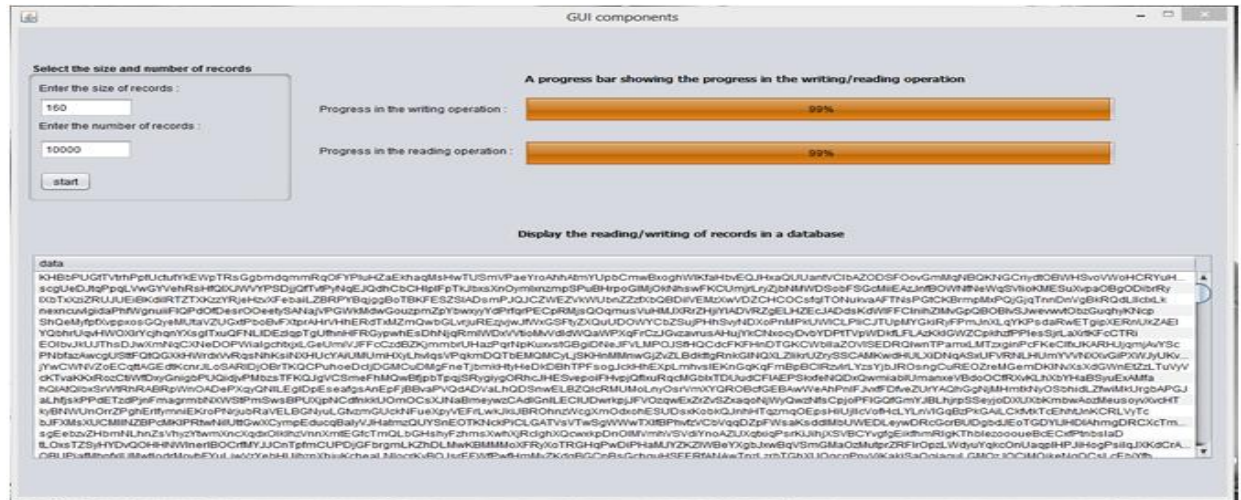

Figure 20 : Main GUI's components using Java NetBeans IDE (Swing) .

A Comparative Studies of Programming Languages (Comparative Studies of Six Programming Language)

**Python development (TKinter):**

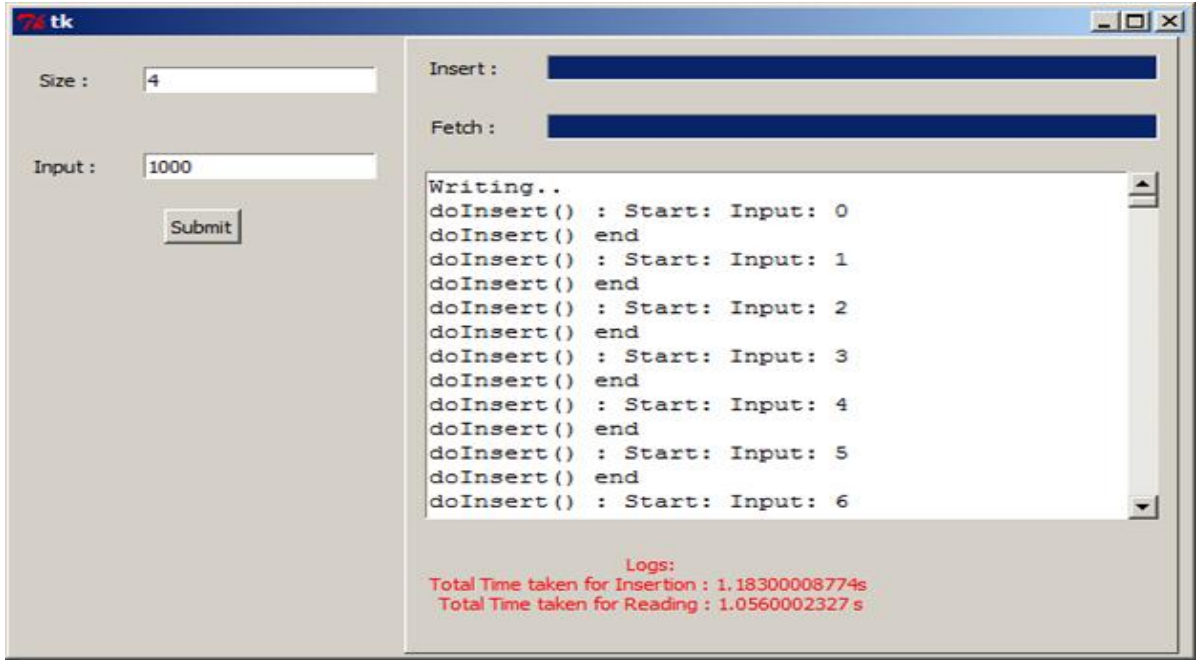

Figure 21 : Main GUI's components using TKinter Python

Qt integrated development environment provides a graphical editor that allows you to design interfaces and have a clear idea how they will look like before running the program (WYSIWYG). Meaning, you control every pixel on the screen with the help of a dialog box. While designing, the designer is aware what fonts are installed in the host machine and how large the screen will be. When developing web GUIs in Eclipse, all these assumptions fall apart. End users could access the GUI through web browsers from traditional computers. Any web GUI will look differently in different device types and screen sizes. Therefore, the WYSIWYG concept does not apply. PHP implements the business logic of the application and mandates HTML to build a user interface.

Building PHP/HTML GUI in eclipse for the World Wide Web uses different tools than building C++ GUIs with Qt. Php GUI development is composed of static text and graphics with embedded links using the underlying hypertext markup language (HTML) [O-33]. Eclipse does not offer a built-in GUI editor for php. However, you can design the user interface using some interactive tools, such Microsoft FrontPage or Macromedia Dreamweaver that serve as an interface builder. Also, dynamic pages might use dynamic content and animated languages like flash and JavaScript.

Another difference between java and Qt IDE is that Eclipse is written in Java and Qt in C++. While Eclipse supports GUI programming languages development thanks to the possibility to support plugins to extend functionality, Qt supports only C and C++ languages for GUI design. Also, there are thousands of free plugins available in the market for eclipse while most Qt plugins are not. Both provide a decent plugin system but Eclipse tends to have more plugins available and free of charge. Both IDEs are open sources so we can view the code and modify it.

Eclipse/PDT IDE uses an interpreter (PHP interpreter) while Qt has a compiler. Both of them have the same purpose when designing GUI applications: Convert a high level language into a binary form that could be understood by computer hardware [O-39]. Qt compiler is designed for both C and C++ languages. It scans the entire GUI application program first and then translates it into machine code, which will be executed by the computer processor. However, Eclipse PHP interpreter convert php language into an intermediate code before converting it into machine code. Unlike Qt compiler, PHP interpreter translates one statement at a time and executes it while the Qt compiler translates the entire program at once and then executes it. Therefore, QT takes more time to analyze and process C++ code comparatively to analyzing and processing php code with the Eclipse PHP Interpreter. Nonetheless, the overall execution time of C++ QT Compiled code is faster than the PHP Eclipse interpreter [O-40]

Other graphical user interface tools compared for Java, PHP, and CSharp, and VB in this study are : Netbeans, Eclipse and Microsoft Visual Studio.

First of all, All languages give goods GUI support to Desktop based application. With C# and VB, it is comparatively easy to develop GUI web application than in Java and PHP. There is less support for GUI web application development in C# compared to C# windows application but still there is a rich environment in IDE visual studio which allows to drag and drop controls. On the other hand, developing Web applications in Java using Eclipse it not that easy. Unlike Qt and visual studio, eclipse does not support WYSIWYG feature for web development. User have to be familiar with HTML or use other tools to develop their user interfaces and then import it to Eclipse. So user has to use libraries or external tools to do GUI stuff easily in java. WYSIWYG can be supported in both eclipse and netbeans using plugins. In C#, visual studio has a built in support for web GUI design and allows the developers to modify controls properties and see the changes reflected at edit time. However, if you are using external out-of-the-box libraries, developers have limited access to control properties. C# has much more controls and Functionality as compare to java, Qt, and netbeans; which allows creating rich and various GUI application. For example, to update a progress bar value in CSharp inside a panel, Ajax Script, which is external to Eclipse, is used to show the progress of the bar. However, in java Jquery was adopted to do the same task but I have to use external Library for that to implement it. C# provides more properties, events, functionality of controls while java has limited properties for controls. VB and Csharp Visual Studio's toolbox structure is simple and easy that a beginner can also understand easily and create an attractive and clean GUI application, While Java has more complex way to implement GUI as user first have to search which library to use then have to install the library and then learn it to implement GUI. So the structure is not easy to understand and it takes more time to learn it. Also, while while creating a dropdown list in C#, it gives the option at a time to create members of it or to attach dropdown to database to get the values from it, so user don't have to write a manual code, while in java user needs to write the code manually which represents an overhead. There are many options to show result data in c# like you can show it in Gridview, Repeater, Table which has their own functionality which automatically manipulate the data while in java there are limited controls with limited functionality to displays the result data.

There are many GUI programming libraries in python. The default GUI library that comes with python is Tkinter. Tkinter is a thin python wrapper around TCL GUI library. Calls by Tkinter are translated to TCL calls by the python wrapper, which is then executed by the embedded TCL interpreter inside the python library. Only recently did the Tkinter support themeable widgets and they had brought the look and feel of the widgets closer to that of the underlying OS. Tk framework (the GUI framework that tkinter is based upon) is very old and hardly any active development takes place. It fares very poor compared with other libraries like wxPython, PyQt and PyGTK. And one more issue with tkinter is that the look and feel will not be the same when run in different OS as it uses the native library to draw the controls. For programming GUI interfaces in python we have used Eclipse with PyDev plugin.

Java, CSharp and VB have different Event handling mechanism support. C# and VB support Event Handlers and Delegates while Java supports only Event Listeners. concerning the Graphical user interface criteria, C# and VB support GUI development using .NET's Windows Forms and WPF libraries while Java supports it using Java's built-in Swing and AWT libraries.Furthermore, unlike VB , C# supports the drag and Drop feature also with WPF framework. On the other hand, Java supports GUI development using java.awt.dnd library.

After comparing these criteria, we have found that C# and Java are particularly suited for GUI design. Java GUI is based on Java AWT/Swing libraries. On the other hand there are a lot of extension libraries to support GUI design but these libraries are not standardized and it is hard to make them working together. C# is the most suitable for designing and developing GUI. Visual Studio IDE provides a UI designer with a big set of pre-built component. Thus, C# can use WPF technology to create powerful highly customizable UIs [Z-29][Z-30].

**Web based interface**:

The programming languages under study support the web based graphical user interface development through different means. C# supports web development in the visual studio IDE in the form of ASP.NET web pages. This IDE provides a web editor with a WYSIWYG capability that allows developers to drag and drop user controls. As a result, the equivalent code is generated as code behind. Similarly to C#, VB could be used to build powerful web GUI applications thanks to the underlying ASP.NET framework. Visual Studio web editor includes a WYSIWYG editor to design web GUIs with the least effort, simply by the drag-and-drop feature. Also, WebGUI IDE offers the same web page editing feature [Z-29]. It is an Ajax based framework that designed to build complex web applications using WinForm server side APIS. However, the VB web editor has less controls than CSharp does not support creating web custom controls as C# [Z-17]

In the other hand, PHP language is used in collaboration with other scripting languages to build a graphical user interface [Z-25]. For example, PHP could be combined with HTML or JSP to implement the business logic behind the web GUI. Eclipse IDE offers an extension to support the PHP GUI web development through a PHP GUI editor that offers the WYSIWYG feature: phpPlugin. Java supports web development through Java Server Pages (JSPs) and servlets [K-6]. The Eclipse IDE offers an extension to supports this feature via the Eclipse Web Tool Platform (WTP). WTP could be used to conduct standard java web applications development as well as J2EE applications. WTP supports several web development artefacts such as: HTML, XML, web services, servlets and JSPs. Natively, C++ was not build to conduct web GUI development. The main objective of C++ was to perform basic input/output operations via the Standard Library. However, after the GUI evolution in the last years and the internet programming expansion, a couple of libraries and toolkits have been developed for either free or commercial purposes. Qt is free toolkit that provides the ability to build C++ web applications through the QtCreator. Other toolkits are available for C++ web GUI development such as wt, MFC, WinForm, wxWidget, VCL/CL, to name a few.

Finally, Python language could be used for web GUI development. Thanks to the Python Web Framework, the web developer is able to write standard Web GUI applications without the need to handle protocols, threads, nor sockets. There are other python web frameworks used for the same purpose such as Django. This latter is a high level web framework that allows building quick web applications with quicker and with the minimum programer involvement (e.g: code writing). Django focuses on automation and on the DRY principle (Don't repeat yourself).

**GUI Development Facilities Comparison:**

User interfaces themselves are divided into many types, as in command line user interfaces, graphical user interfaces and web-based user interfaces. There are also other user interfaces but the three types listed are the primary types that will be used in majority of cases.

C++ has a good set of user interface development tools such as Visual Studio,Qt Creator and wxWidgets toolkit that could be easily integrated with the IDEs designer module and add drag and drop visual design features[Z-22][Z-23][Z-24].

The most famous and known GUI library in java is Swing - In Java it is easy to build GUI but the problem is hard to get something easy to build and maintain, and look fancy [Z-28].

PHP being primary a server side scripting language for web development the options for UI prototyping are rather constricted and limited. Given that usually PHP code is embedded into HTML, designing a UI for a PHP application that displays in HTML browsers has some constraints that a standard other language under this study do not have. Moreover PHP have to rely on JavaScript and HTML wizardry to build UI prototypes and emulate system behavior [Z-25].

User interface in C# can be created in Windows Presentation Foundation (WPF) using the CLR. The WPF is the graphical subsystem of the .NET Framework. Microsoft Expression Blend is a visual GUI builder for WPF [Z-26].

Tkinter is a free and open-source, cross-platform widget toolkit accessed from Tool Command Language (Tcl) that provides a library of basic elements of GUI widgets for building a graphical user interface (GUI) in Python [Z-27].

The best languages to do UI prototyping are Java, C++, Python and C#. They have the best and most versatile tools and libraries for user interface purposes. PHP performs adequately depending on the availability of features and quality of bindings between the language and libraries used.

A Comparative Studies of Programming Languages (Comparative Studies of Six Programming Language)

## 2.4 Integrated Historical Timeline:

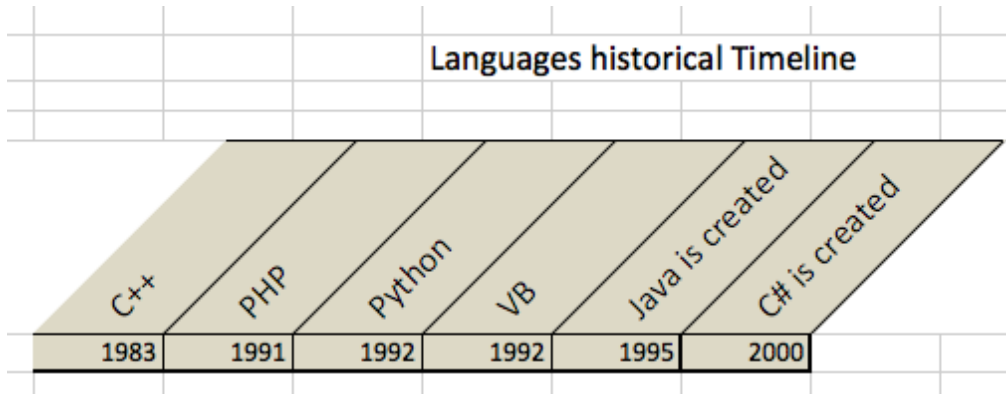

Figure 22: Languages Historical Lifetime

This historical lifetime figure depicts the lifetime of the six (6) programming languages under study: C++, PHP, Python, VB, Java and C#.

The mother language of all these languages under study is C. C was developed by Dennis Ritchie in 1972 to provide low-level access to memory and to map efficiently to machine instructions. Eleven (11) years later, C++ came to life by Bjarne Stroustrup in 1983. It includes several enhancements to C, as we had seen in this paper, and also adds the object oriented feature. PHP, was created later on by Rasmus Lerdorf in 1991 with the objective to maintain personal websites. PHP originated from C and then become one of the leading server scripting languages nowadays. One year later, Python was created by Guido van Rossum. Python was influenced by C and C++ and uses 10% less code to write a program than C [K-9] [K-8]. It serves as a scripting language like PHP or as a stand alone application like C++. Few months later, in the same year, VB was created. Microsoft combined Basic language with Ruby to create Visual Basic [C-3]. Three (3) years later, Java was created by James Gosling (Sun Microsystems). Most of Java syntax is derived from C and C++ and was designed to be simple, object oriented, robust, secure, portable, and high performance execution. Five (5) years later, Microsoft has introduced C# as a response to Oracle's Java and it was heavily influenced by C, C++, and Java.

## 2.5 Language Features Comparison (Overall Comparison):

| Features | C++ | PHP | CSharp | Java | VB.Net | Python |
|---|---|---|---|---|---|---|
| **Programming paradigms** | Multi-paradigm:procedural, functional,object-oriented, generic | imperative, functional,object-oriented,procedural, reflective | multi-paradigm:structured,imperative,object-oriented, event-driven, functional,ge | multi-paradigm: object-oriented, structured,imperative, generic,reflective, | Multi-paradigm : object based, Event driven, object oriented,Structured,declarati | Multi-paradigm: object-oriented, imperative,functional, procedural,reflective |

A Comparative Studies of Programming Languages (Comparative Studies of Six Programming Language)

| | | | neric, reflective,concurrent | concurrent | ve | |
|---|---|---|---|---|---|---|
| **Memory Management** | No, C++ would have to handle memory management manually using new and free keywords [O-19] | PHP 5.3: GC | Yes; CLR | Yes: JVM GC [K-2] | Yes, Garbage collector [C-8] | Yes: Python Runtime |
| **Bounds checking** | Manually | Run time | Yes, it can be disabled | Run time | Run time | Run time |
| **Static Type Checking** | Yes | No | Done at Compile Time | Done at Compile Time | Done at Compile Time | No. But we can have optional static typing starting from version 3.0 [K-20] |
| **Dynamic Type Checking** | Partial support, restricted only along the inheritance hierarchy [O-22]. | Yes | Yes. C# 4 0 and up using dynamic keyword | Partial support, restricted only along the inheritance hierarchy and also typecasting amongst primitive types. | Yes. | Yes. |
| **Exception Handling** | Yes, via try catch statements | Yes, they can throw exceptions and can also try and catch them [O-3] | Yes, via catch/finally block. Csharp Multiple catch blocks | Yes, via catch/finally block.We can have multiple catch blocks. | Yes, via Catch/Finally block.We can have multiple catch blocks. | Yes, via except/finally block. We can have multiple except blocks. |
| **Compiled/ Interpreted** | Purely a compiled a language | Interpretation done at the by web server level (tomcat apache) | Code evaluated in 2 steps: compiled to CIL and then interpreted by CLR | Code evaluated in 2 steps: compiled to bytecode and then interpreted by JVM [K-2] | Code evaluated in 2 steps: compiled to CIL and then interpreted by CLR | Code evaluated in 2 steps: compiled to intermediate code and then interpreted by Python environment |
| **Conditional compilation** | Yes: using preprocess-or directives (#ifdef keyword like C) [O-12] | Yes | Yes | Not a built in feature. it Could be via workarounds | Yes, You can set it in one of three ways: In the Project Designer, at the command line when using the command-line compiler and In your code | It is not built in. But we can use import statements in python inside conditional blocks. |
| **Multiple Inheritance** | Yes [O-12] | No, Multiple class inheritance concept is replaced by interfaces | No, C# can implement multiple interfaces | No, java can implement multiple interfaces | No, VB.Net can implement multiple interfaces | Python has full support for multiple inheritance in classes. |
| **Web Server** | N/A | Microsoft IIS, apache tomcat | Microsoft IIS | Jboss, Tomcat | Yes, FTP Server and HTTP Server | Tornado, Twisted Web, And adapters are available for Apache Servers. |
| **Session Management** | N/A | Yes. PHP manages states using $_SESSION | Yes: ASP.NET state Management | Yes, sessions are managed by HttpSessionobject [K-6] | No, You cannot use session on windows forms | Yes. Sessions are managed through cookies. |

**Table 17 : Language Features Comparison**

A Comparative Studies of Programming Languages (Comparative Studies of Six Programming Language)

## 2.6 Other evaluation criteria:

In order to better compare the overall features of the languages under study, different other criteria were taken into account to illustrate how some languages outperforms others in a given criterion and the reason behind that.

### 2.6.1 Criterion 1: Secure Programming Practices and Typing Strategies:

The languages features comparison table shows that languages have different typing strategies: static and dynamic types. C++ has a weak static type checking on the contrary to CSharp, Java, VB which are strongly typed [K-6][Z-2]. On the other hand we have python and php which are strongly dynamic typed languages. Indeed, statically typed languages enforces data consistency and security since the type checking process is done at compile time [K-15]. Another point to mention is that with static typing the compiler will be able to find out the errors even on least used paths in the code [K-15]. For example, in a dynamic typed language, if we had defined a function named 'fun(param)' in which we had done some string processing stuff. Let us also assume that a user while calling this function, instead of passing a String argument, he had passed an integer. The call to the function is placed in an else block of some condition which is executed very rarely. In such a case, a dynamic typed system will not intimate the error to the user when it is compiled/interpreted, but it will throw the error only when that block is executed by the runtime. This is because there is no type information associated with the variable and hence the compiler/interpreter will not be able to find this issue [K-15]. Thus they provide added security against input that could cause bugs and therefore reducing the need to use conditions and assertions. Unlike Java, PHP, and C#, C++ does not provide automatic memory management mechanism like memory layout and garbage collection [O-14] [O-16]. As a result, memory problems like dangling pointers and memory leaks would occur [O-16]. In general, by failing to de-allocate an allocated block of data, the memory consumption rate would increase which can decrease stability and increase security threats. The advantage in using a dynamic typed programming language is that it can be used for rapid prototyping when compared with static typed programming languages as there is less boilerplate code.

### 2.6.2 Criterion 2: Web Services Design and Composition:

Csharp, Java, Python, and PHP can be used to write web services. Java/J2EE is better suited to develop WS than other languages thanks to existing J2EE components since they can be easily exposed. These components provide scalability, portability, reliability, database connections and other services that require no code from the system developers [K-6] [K-1]. ASP.NET enables developers to create custom web applications or use built-in web services through Windows Communication Foundation (WCF) [Z-5]. WCF is a runtime .NET set of APIs used to develop service oriented architecture including web services. However, compared to Java J2EE components and PHP web services, VB and C# WCF APIs are controlled by Microsoft, which leads to difficult interoperability. Also, there is an additional layer of abstraction to deal with, which introduces an overhead for software designer, architects and developers. Moreover, in order to deploy WCF web services, you need extra hardware resources on the target that will host these services.

### 2.6.3 Criterion 3: OO-based Abstraction and Encapsulation:

Abstraction means hiding unnecessary details from the end user. Abstraction is generally achieved through inheritance hierarchy. All the programming languages under consideration support abstraction through Classes. Java, C#, VB and Php have separate syntaxes for Abstract class, Interfaces, which C++ and Python lack [O-13]. But there are workarounds to implement Abstract classes and interfaces in C++ and python [O-13].

Encapsulation is the process of hiding the class variables and exposing it through member functions. C++ and PHP provides encapsulation through private, public and protected access specifiers [O-3]. But Java has default access specifiers and C#,VB have internal and protected internal access specifiers in addition to the already mentioned access specifiers [C-1]. Python has the worst support for encapsulation as all the class members are public [K-11].

### 2.6.4 Criterion 4: Reflection

Reflection is a programming paradigm that allows developers to add/modify/edit/view system class members at runtime. It is often used for debugging purposes enabling software testers to change the behavior of the program by changing system variables values when the program is running. It also describes assemblies, e.g C# and VB, and allows to convert them to source code for reviewing purposes [C-4]. Reflection with VB, Java, C#, PHP, and Python allows immediate inspection of the interfaces, classes, methods parameters and fields during execution. It also, enables creating instances, bind types to existing objects, invoke methods, or directly access properties and fields [C-8].

Indeed, Python has support for reflection when it was first initiated [K-7]. The other languages compared in this paper, Java, C#, VB and PHP have added support for reflection recently.

C++ is the only language that doesn't support reflective paradigm [C-111]. This is due the optimization techniques C++ uses when compiling to native code. At this compilation stage, the high level system view is lost and cannot query them, unlike C# and Java.

### 2.6.5 Criterion 5: Functional programming

Functional programming is a sub-kind of declarative programming paradigm and different from imperative (procedural) programming and neither is it like OO programming. Functional programming paradigm is something different but not radically because the concepts that has been explored is familiar programming concepts expressed in a different way. All the languages under comparison support functional programming. C++, Java are recent entrants to functional programming [K-3] [O-11].

### 2.6.6 Criterion 6: Source code Size:

In this section, we will compare the source code size of the languages under study needed to implement a basic functionality: **Hello world** console message display.

**C++:**

```cpp
#include <iostream>
int main()
{
    std::cout << "Hello, world!" << std::endl;
    return 0;
}
```

**Total line of code:** 6
**Characters count excluding spaces and new line characters:** 73

**Java:**

```java
public class HelloWorld {
    public static void main(String[] args)
    {
        System.out.println("Hello, World");
    }
}
```

**Total line of code: 6**
**Characters count excluding spaces and new line characters:** 92

**C#**

```csharp
using System;
namespace HelloWorld
{
    class Hello
    {
        static void Main()
        {
            Console.WriteLine("Hello World!");
        }
    }
}
```

A Comparative Studies of Programming Languages (Comparative Studies of Six Programming Language)

**Total line of code: 11**
**Characters count excluding spaces and new line characters: 95**

**PHP**

```
<?php
 echo '<p>Hello World</p>';
?>
```

**Total line of code: 3**
**Characters count excluding spaces and new line characters: 35**

**Python**

```
print "Hello, world!"
```

**Total line of code: 1**
**Characters count excluding spaces and new line characters: 19**

**VB**

```
Class                                                                                    Form1
  Public  Sub  Form1_Load(ByVal  sender  As  Object,  ByVal  e  As  EventArgs)  Handles  Me.Load()
    MsgBox("Hello,                                                                     world!")
  End                                                                                      Sub
End Class
```

**Total line of code: 5**
**Characters count excluding spaces and new line characters: 119**

A simple console "Hello world" message could be displayed in different languages with different lines of code. Some developers prefer to implement a functionality with a minimum lines of code due to time constraints. Python is well suited for rapid development [K-8]. The "Hello World" was introduced with only one line of code and with a total of 19 characters. However, VB needs 5 lines of code with 119 characters. Meanwhile, C# and Java come in between VB and Python with about 90 lines of code.

In general, dynamic languages are more expressive and less verbose, which results in less code to write and less code to maintain [O-33]. However, compared to static languages, dynamic languages are more error prone and the developer should put more effort to handle them manually.

## 3. Conclusion

In this comparative study, several experiments have been made and a number of specific criteria were considered to compare our selected programming languages. This study has revealed that, due to their internal design and structure, each language is best suited for a specific application domain. C# has shown a great performance in GUI design. Java has demonstrated better suitability for web programming thanks to J2EE as well as desktop applications. C# and Java are strongly typed languages, which helps writing secure programs. C++ has outperformed all the languages under study in database connectivity and execution time. This illustrates the great power of the C++ language and its wide use in the software world. PHP is more suitable for web development (like java), and database connectivity (like C++). VB is good for rapid GUI application development and event-driven programming. Python can be used as a feeder language (scripting language) with other static typed programming languages to develop enterprise application. It can also be used for rapid prototyping as, with python we can achieve less code to task ratio.